האוניברסיטה העברית בירושלים
The Hebrew University of Jerusalem

Faculty of Mathematics and Natural Sciences

The Racah Institute of Physics

# Quantum Langevin Dynamics

M.Sc. Thesis

By

Mohammed Attrash

Under the supervision of

Prof. Roi Baer

Acknowledgements

Thanks for Prof. Roi Baer for his patient and guidance in this project, I study many things from him.

Thanks also to my parents whom stand with me.

## ABSTRACT

Previous year's researchers began to simulate open quantum system, taking into account the interaction between system and the environment.

One approach to deal with this problem is to use the density matrix within the Liouville-von-Neumann formalism or the Markovian variant – the Lindblad equations. Another way is to use a stochastic approach where a random force is added to the system.

The benefit of the stochastic approach is to solve the dynamics of the system with less time and memory than the density matrix approaches.

In this project we want to develop a stochastic approach that can deal with the stochastic wave functions approach. We did this on a 2-level system and found that it works well when comparing to a density matrix approach.

Next, we tested a quantum particle connect to a bath of harmonic oscillators using the stochastic approach. We found that a friction term is necessary and applied it. Like in the classical Langevin equations the friction constant and the random force fluctuations are related by the fluctuation-dissipation constant. We showed that with friction the dynamics decays to an ensemble with energy of $E_{gs} + k_B T$.

However, we also found there are problems. The system seems to absorb energy indefinitely if the temperature is higher than the zero point energy or if the system is a Morse oscillator. Thus more research is required to make this method work.

# 1 Table of Contents

# 1. Introduction

For many problems we need methods that can take into account the open-system nature of the problem. Such approaches typically use the density matrix. This is a huge practical problem when the system is very large, as the case, for example, in Nanocrystals, because storing and manipulating the DM is beyond the capability of present day technology.

## 1.1 Open systems in classical mechanics: Langevin Dynamics

In classical physics the description of open systems uses Langevin dynamics. This approach introduces a random force $F_R$ that describes the effect of a bath and supplements it with a friction force to dissipate away the energy. The Langevin equation is thus of the following form [1]:

$$dp(t) = f\big(q(t)\big)dt - \gamma p(t)dt + \sqrt{2m\gamma k_B T}dW(t)$$

Where $p$ is the momentum of the system, $m$ the mass and $\gamma$ is the friction coefficient, $f(q) = -\frac{\partial U}{\partial q}$ is the deterministic force ($U$ a potential) and $dW(t)$ is a Wiener noise zero memory autocorrelation: $\langle dW(t)dW(t')\rangle = \delta(t-t')$. It can be shown [2] that any observable averaged over a long time trajectory will yield its thermal average. For example, in one dimension the kinetic energy $\left\langle \frac{p^2}{2m} \right\rangle$ will average to $\frac{1}{2}k_B T$.

Langevin dynamics is routinely used in a huge variety of applications where thermal averages are needed, but it is also used often when a system is driven out of equilibrium by an external force.

An analogous approach to the Langevin dynamics in quantum mechanics is the subject of this thesis

## 1.2 The density matrix as an open system state

Suppose we are not certain what the state of the system is. This can be caused when a system is coupled to a heat bath and we have no control over the precise state of the bath. We have a set of orthonormal states $\psi_n(x)$ and each of these is a possible state with probability $0 \le w_n \le 1$ and $\sum_n w_n = 1$:

$$\rho(x,x') = \sum_n w_n \psi_n(x')^* \psi_n(x)$$

$$\hat{\rho} = \sum_n w_n |\psi_n\rangle\langle\psi_n|$$

$$tr[\hat{\rho}] = \sum_n \langle\psi_n|\hat{\rho}|\psi_n\rangle = \sum_n w_n = 1$$

A pure state is when there is no uncertainty, i.e. $w_1 = 1$ and then $\hat{\rho} = |\psi\rangle\langle\psi|$ is projection operator, so $\hat{\rho}^2 = \hat{\rho}$ and $tr[\rho^2] = \sum_n w_n^2 = w_1^2 = 1$. In mixed state, $\hat{\rho}^2 \neq \hat{\rho}$ and $tr[\rho^2] = \sum_n w_n^2 < 1$.

For a closed system the dynamics of the DM is derived from the Schrödinger equation for each of the $\psi_n$:

$$i\dot{\hat{\rho}} = i\sum_n w_n|\dot{\psi}_n\rangle\langle\psi_n| - i\sum_n w_n|\psi_n\rangle\langle\dot{\psi}_n| = \hat{H}\sum_n w_n|\psi_n\rangle\langle\psi_n| - \sum_n w_n|\psi_n\rangle\langle\psi_n|\hat{H}$$
$$= \hat{H}\hat{\rho} - \hat{\rho}\hat{H} = [\hat{H},\hat{\rho}].$$

An open system is usually obtained when we consider a large system composed of a "system" and a "bath". The total Hamiltonian is then:

$$H = H_0 + H_B + V_{SB}$$

Where $H_0$ is the system Hamiltonian and $H_B$ is the bath Hamiltonian. $V_{SB}$ is the coupling between system and bath, which is often macroscopic in nature.

It is often a good approximation to assume that the bath is in thermal equilibrium:

$$\hat{\rho}_B = \frac{e^{-\beta\hat{H}_B}}{Z_B}$$

Where $Z_B = tr[e^{-\beta\hat{H}_B}]$ and the temperature $T = \frac{1}{k_B\beta}$. $W$e then define a reduced DM:

$$\hat{\sigma} = tr_B\hat{\rho}$$

where the trace is over bath degrees of freedom. The dynamics of the system is linear in the reduced DM $\hat{\sigma}$ so it is determined by a non-Markovian equation of the form

$$\dot{\hat{\sigma}}(t) = \int_{-\infty}^{t} \hat{\hat{L}}(t,\tau)\hat{\sigma}(\tau)d\tau$$

## 1.3 Lindblad (DM formalism)

Lindblad determined to simplify the equation of motion for the reduced DM $\hat{\sigma}$ studied a *Markovian* (no memory) form, $\hat{\hat{L}}(t,\tau) = \mathcal{L}\delta(\tau)$:

$$\dot{\hat{\sigma}}(t) = \mathcal{L}\hat{\sigma}(t)$$

He demands certain conditions must hold:

1) $tr[\hat{\sigma}] = 1$
2) $\hat{\sigma}$ is positive definite: for any $v \neq 0$ we must have $\langle v|\hat{\sigma}|v\rangle > 0$.
3) Translational invariance
4) Asymptotic approach to equilibrium

He then found a canonical form for "Markovian Liouville equations".

An example of such a "Lindblad" form is the following equation of motion for the reduced DM $\hat{\sigma}$:

$$\dot{\hat{\sigma}}(t) = \mathcal{L}\hat{\sigma}(t) = -i\left[\hat{H}_0, \hat{\sigma}(t)\right] + L\hat{\sigma}(t)L^\dagger - \frac{1}{2}\{L^\dagger L, \hat{\sigma}(t)\}$$

If $L^\dagger = L$ is hermitean we have a simpler form:

$$\dot{\hat{\sigma}}(t) = -i\left[\hat{H}_0, \hat{\sigma}\right] - \frac{1}{2}\left[\hat{L}, \left[\hat{L}, \hat{\sigma}\right]\right]$$

This is because:

$$\left[\hat{L}, \left[\hat{L}, \hat{\sigma}\right]\right] = L\left[\hat{L}, \hat{\sigma}\right] - \left[\hat{L}, \hat{\sigma}\right]L = L^2\sigma - 2L\hat{\sigma}L + \sigma L^2 \tag{1}$$

One can use this equation to determine the dynamics of open systems if the operators, characterizing the system-bath coupling and the state of the bath are known. However, such a scheme is limited to small systems because $\hat{\sigma}$ is usually a density matrix on the system Hilbert space.

According to [3] there are several choices for dissipative operators, and for the case of pure dephasing when the bath cause decoherence in the system but no energy is exchanged the Lindblad operators are diagonal in the basis of the system eigenstates, like the case we take the Hamiltonian itself as Lindblad operators.

## 1.4 Stochastic approach to Lindblad formalism

In appendix 3 we show that formally, instead of solving the Lindblad equations for the reduced DM we can use wave functions obeying the Schrödinger equation supplemented with a time-dependent random force $\hat{F}$:

$$i\hbar\dot{\psi}(t) = \hat{H}_0\psi(t) + \hat{F}(t)\psi$$

$\hat{F}(t)$ is designed in such a way that the reduced DM is obtained from averaging over the individual DMs: [4]

$$\hat{\sigma}(t) = \langle \ |\psi(t)\rangle\langle\psi(t)| \ \rangle.$$

The relation between the characteristics of $\hat{F}$ and the constant $D$In chapter 2 we show an example where such an approach works very well, and the dynamics of a reduced DM of a 2-level system coupled to a bath can be described.

However for a larger system we are missing a friction (term similar to the one in classical Langevin dynamics) and this causes problems of stability. We will study ways to remedy this problem.

One of the benefits of this approach when we dealing with large systems like nanoparticles in 3D, our method uses much less memory and can expedite computational speed.

# 2 Stochastic Schrödinger equation approach for the 2-level system

In this section we study the application of stochastic wavefunctions as a method for solving the Lindblad equations in the simplest possible context, namely for a two-level system. For such a simple system a numerically exact density matrix solution to the Lindblad equation exists and thus we can compare to "numerically exact results".

## 2.1 The 2-level system used for benchmarking the stochastic approach

We have a Hamiltonian which describes standard model of TLSs [5]:

$$H_0 = \frac{\epsilon}{2}\sigma_z$$

Where $\sigma_z = \begin{pmatrix} 1 & 0 \\ 0 & -1 \end{pmatrix}$ (a Pauli operator). Clearly the eigenvalues are $\epsilon_{\pm}^{(0)} = \pm\frac{1}{2}\epsilon$. Under a perturbation which couples the two levels, $\Delta V = \frac{\Delta}{2}\sigma_x$ where $\sigma_x = \begin{pmatrix} 0 & 1 \\ 1 & 0 \end{pmatrix}$, the Hamiltonian becomes:

$$H_0 = \frac{\epsilon}{2}\sigma_z + \frac{\Delta}{2}\sigma_x$$

The eigenvalues of the Hamiltonian are:

$$\epsilon_{\pm} = \pm\frac{1}{2}\sqrt{\epsilon^2 + \Delta^2}$$

The two-level system is often an approximation to a more complicated system, for example, the system of two coupled wells shown in the following figure:

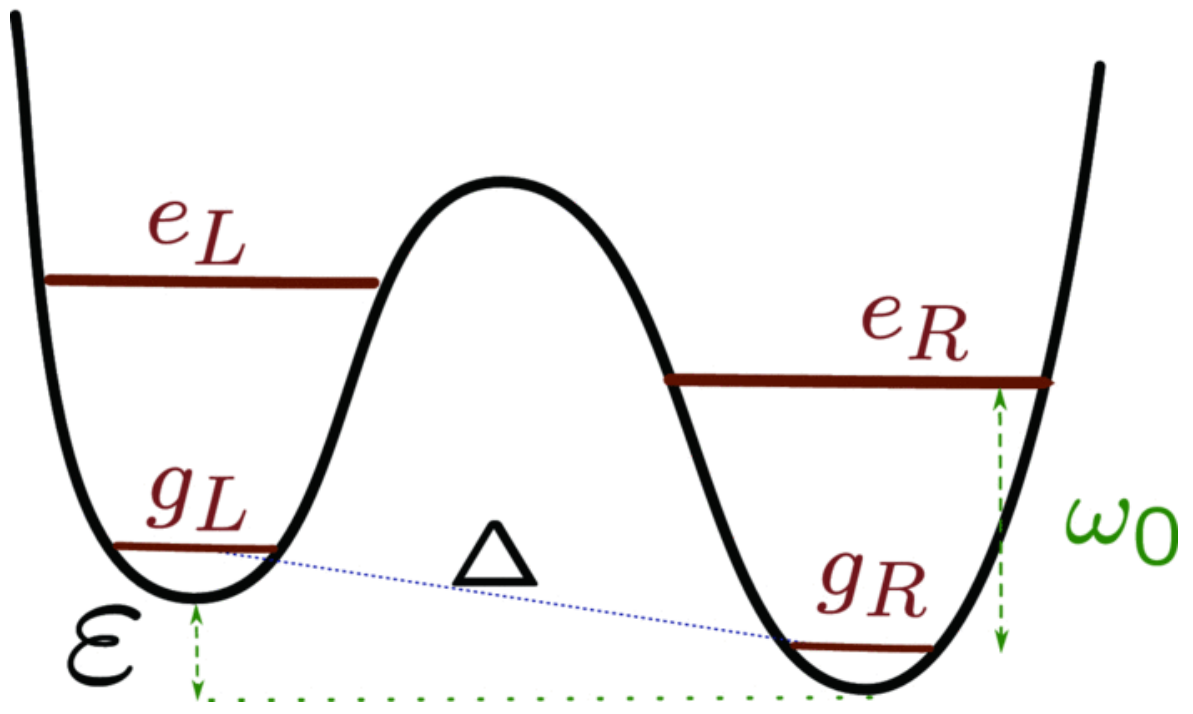

The potential $V$ is composed of two weakly coupled left and right wells. If the coupling is neglected then we can define the ground state and excited state energies on the left, $g_L$ and $e_L$ respectively, and the ground state wave function $\psi_L$. Similar quantities can be defined for the right well. The two-level approximation is valid if $e_i - g_i \gg \epsilon, \Delta\ (i = L, R)$ where $\epsilon = |g_L - g_R|$ and $\Delta = \langle\psi_L|V|\psi_R\rangle$.

One limit often treated is the weak coupling limit, $\Delta \ll \epsilon$, where the new eigenvalues are similar to the unperturbed ones

$$\epsilon_{\pm} = \pm\frac{\epsilon}{2}\sqrt{1 + \left(\frac{\Delta}{\epsilon}\right)^2} \approx \pm\frac{\epsilon}{2}\left(1 + \frac{1}{2}\left(\frac{\Delta}{\epsilon}\right)^2\right)$$

However we will study a strong coupling case where $\Delta \approx \epsilon$ in the sudden approximate where the initial state at $t = 0$ is the ground state of the unperturbed system while the Hamiltonian which performs the time propagation is the full perturbed Hamiltonian. Our focus will be the connection to a heat bath which we will describe using the density matrix:

$$\hat{\rho} = \rho_{00}|0\rangle\langle 0| + \rho_{01}|0\rangle\langle 1| + \rho_{10}|1\rangle\langle 0| + \rho_{11}|1\rangle\langle 1| = \begin{pmatrix}\rho_{00} & \rho_{01}\\ \rho_{10} & \rho_{11}\end{pmatrix}.$$

Obeying the Lindblad equations:

$$\dot{\hat{\rho}} = -\frac{i}{\hbar}[\hat{H}_0, \hat{\rho}] - \frac{D}{2\hbar^2}[\hat{x}, [\hat{x}, \hat{\rho}]] \qquad (2)$$

In the Lindblad equation $\hat{x}$ can in principle be any operator. In the present case we take

$$\hat{x} = \sigma_z$$

## 2.2 Stochastic Schrödinger equation approach

Stochastic Schrödinger equation approaches attempt to replace the Lindblad equation by a time-dependent Schrödinger equation with random potentials. The usual procedure uses non-Hermitean and non-linear Schrödinger equations which are constructed in a way that perseveres populations [6].

In this work we examine a different route, namely one which is fully Hermitian, although nonlinear.

### 2.2.1 Hermitian stochastic Schrödinger dynamics

For the case of a two-level system the Lindblad equation is replaced by a time-dependent Schrödinger equation with random potentials:

$$i\hbar\dot{\psi} = \hat{H}(t)\psi. \tag{3}$$

Where the stochastic-hermitian Hamiltonian is:

$$\hat{H}(t) = \hat{H}_0 + f(t)\hat{x} \tag{4}$$

and $\hat{H}_0 = \frac{\epsilon}{2}\sigma_z + \frac{\Delta}{2}\sigma_x$ while $\hat{x} = \sigma_z$ or $\sigma_x$ (in the examples below we take $\hat{x} = \sigma_z$ for definiteness). Where $f(t)$ is a random white-noise field of amplitude $\sigma_F$:

$$\begin{gathered}\langle f(t)\rangle = 0 \\ \langle f(t)f(t')\rangle = \sigma_F^2\delta(t-t').\end{gathered} \tag{5}$$

The connection between the wave functions propagated by the stochastic wave function and the corresponding Lindblad equation (2) is three fold:

1) The amplitude of the random force obeys: $2\sigma_f^2 = D$ where $D$ is the coefficient of the Lindblad term in Eq. (2).
2) The density matrix is obtained as an average over 1-particle pure-state density matrices obtained in each separate realization of the stochastic dynamics: $\hat{\rho}(t) = \langle \ |\psi(t)\rangle\langle\psi(t)| \ \rangle$. Note that since $|\psi(t)\rangle\langle\psi(t)|$ is positive definite (i.e. for any ket $|v\rangle \neq 0$ we have: $\langle v|\psi(t)\rangle\langle\psi(t)|v\rangle > 0$) its average is also positive definite (the sum of positive definite operators is also positive definite).
3) Assuming we start with a pure state we simply propagate this state according to the stochastic Schrödinger equation. A generalization to mixed states is also possible (see section 2.4.3).

### 2.2.2 The numerical solution to the stochastic Schrödinger equation

We began from the ground state of unperturbed Hamiltonian $\psi_0 = \begin{pmatrix}0\\1\end{pmatrix}$, we want to propagate $\psi$ in time, so we use:

$$\psi(t = N\delta t)_{\delta t\to 0} = \prod_j^N U_j(\delta t)\,\psi_0$$

Where:

$$U_j(\delta t) = e^{-iH\delta t}$$

In our case:

$$U_i(\Delta t) \approx e^{-\frac{i\frac{\Delta}{2}\sigma_x\delta t}{2}} e^{-i(\frac{\epsilon\sigma_z}{2}+f)\delta t} e^{-\frac{i\frac{\Delta}{2}\sigma_x\delta t}{2}}$$

We use the fact:

$$e^{-i(\sigma_z+f)\Delta t} = \begin{pmatrix} e^{-i\left(\frac{\epsilon}{2}+f\right)\delta t} & 0 \\ 0 & e^{i\left(\frac{\epsilon}{2}+f\right)\delta t} \end{pmatrix}$$

$$e^{-\frac{i\Delta\sigma_x\delta t}{4}} = \cos\left(\frac{\Delta\delta t}{4}\right)\hat{I} + i\sin\left(\frac{\Delta\delta t}{4}\right)\hat{\sigma}_x = \begin{pmatrix} \cos\left(\frac{\Delta\delta t}{4}\right) & i\sin\left(\frac{\Delta\delta t}{4}\right) \\ i\sin\left(\frac{\Delta\delta t}{4}\right) & \cos\left(\frac{\Delta\delta t}{4}\right) \end{pmatrix}$$

Where $\Delta$ come from the Hamiltonian.

## 2.3 The numerically exact solution to the Lindblad equation

We use the formalism described in references [7] and [8] and define three operators:

$$\rho_x = \frac{1}{2}(\rho_{01}+\rho_{10}),\, \rho_y = \frac{i}{2}(\rho_{01}-\rho_{10}),\, \rho_z = \frac{1}{2}(\rho_{00}-\rho_{11})$$

Treating the density matrix as a 3-vector:

$$\rho = \begin{pmatrix} \rho_x \\ \rho_y \\ \rho_z \end{pmatrix}$$

So we can write this equation as:

$$\dot{\rho} = iL\rho$$

Where

$$L = \begin{pmatrix} -D & -\epsilon & 0 \\ \epsilon & -D & \Delta \\ 0 & -\Delta & 0 \end{pmatrix}$$

This equation is of the optical Bloch equation type [9]. In order to obtain the numerically exact solution to the Lindblad equation we propagate in time, where the initial state $\rho_0$ is taken as $\rho(t=0)$. We take smack time steps $\Delta t$ and propagate in these steps as follows $t_n = n\Delta t$:

$$\rho(t_n) = (1 - iL\Delta t)^n \rho_0$$

This method is not very efficient because each step is only second order. However, dues to the simplicity of the problem, we can afford to make small steps and converge the results to full numerical accuracy.

## 2.4 Numerical comparison: stochastic Schrödinger vs. Lindblad Equations for 2-level systems

Except noted otherwise, in all cases studied, we start from the unperturbed ground state $\rho_{00} = 1, \rho_{11} = 0$ and describe the evolution of the populations and coherences in the DM. We show several cases for the 3 parameters of the model ($\epsilon$, $\Delta$, and $D$) the populations and coherences as a function of time calculated using the numerically exact propagation and compared to the stochastic Schrödinger equation approach.

### 2.4.1 Case of equal parameters $\epsilon = \Delta = 0.2$

#### 2.4.1.1 Zero coupling to bath: $D = 0$.

The time propagation shows Rabi oscillations (see Appendix 2 for discussion of Rabi oscillations) in the population (see left panel below) between the sites with frequency $\Omega_R = \sqrt{\epsilon^2 + \Delta^2} = \sqrt{0.2^2 + 0.2^2} = 0.28$ and period 22. The coherences shown on the right panel do not decay. Here there is no random field and the results of solving the Lindblad and the Schrödinger equation lead to identical populations and coherences, as they must.

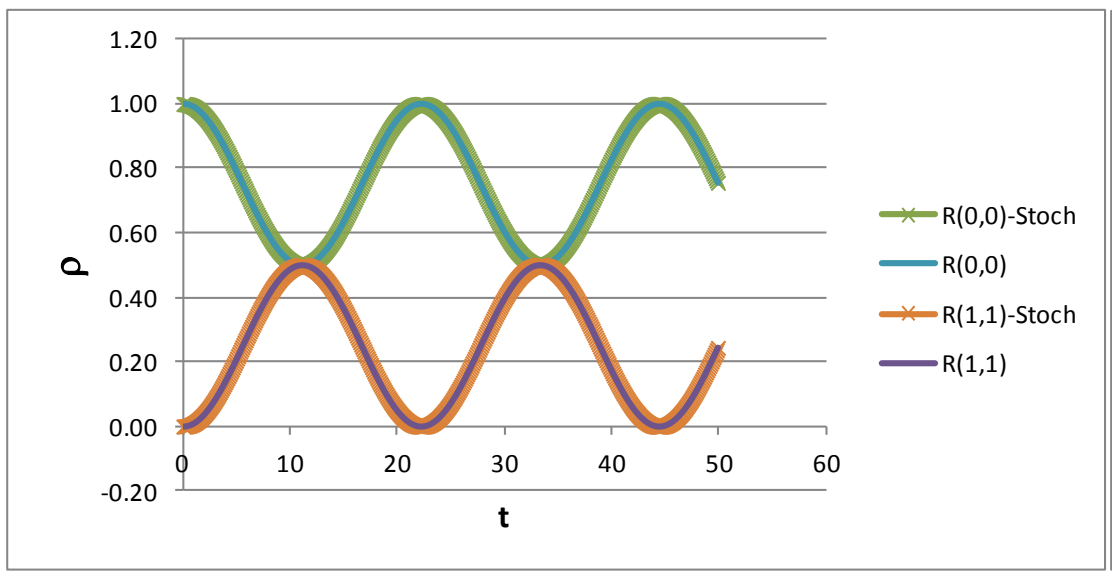

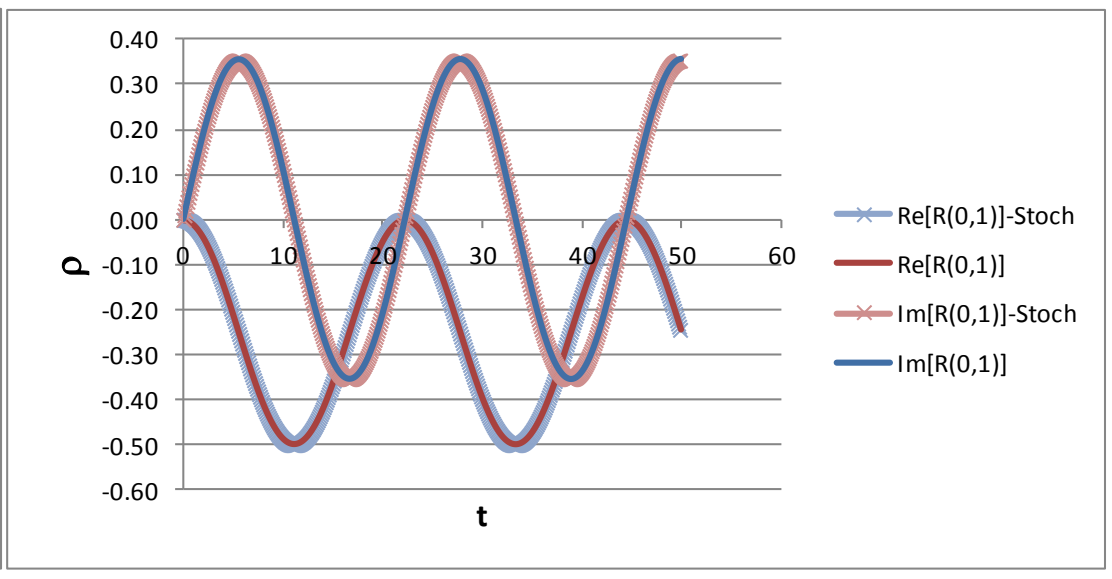

**Figure 2.1: The DM populations (left) and coherences (right) for zero coupling. Averaged dynamics based on the stochastic Schrödinger equation are shown together with the numerically exact Lindblad dynamics. With $\epsilon = \Delta = 0.2$ and zero coupling $D = 0$.**

### 2.4.1.2 Small coupling to bath: $D = 0.03$.

Here the system displays damped population oscillations that decay to 0.5 with a single exponential having a time constant of $\lambda = 0.016$ Coherences decay to zero with the same time constant.

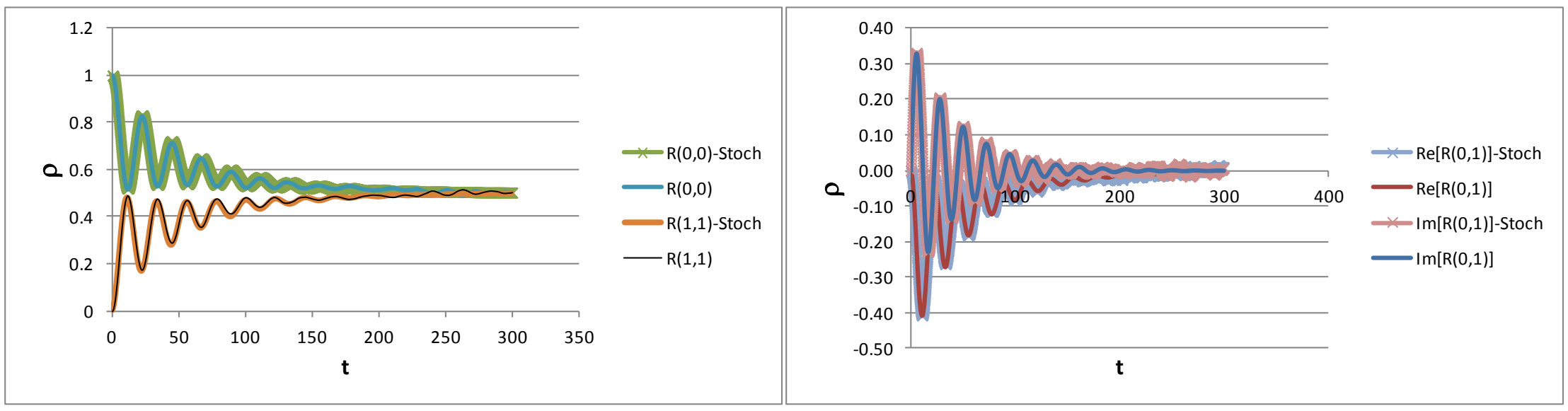

Figure 2.2: Same as Figure 2.1 but for $D = 0.03$.

### 2.4.1.3 Medium coupling to bath: $D = 0.1$.

Here the system displays damped population oscillations that decay to 0.5 with a single exponential having a time constant of $\lambda = 0.05$ Coherences decay to zero with the same time constant.

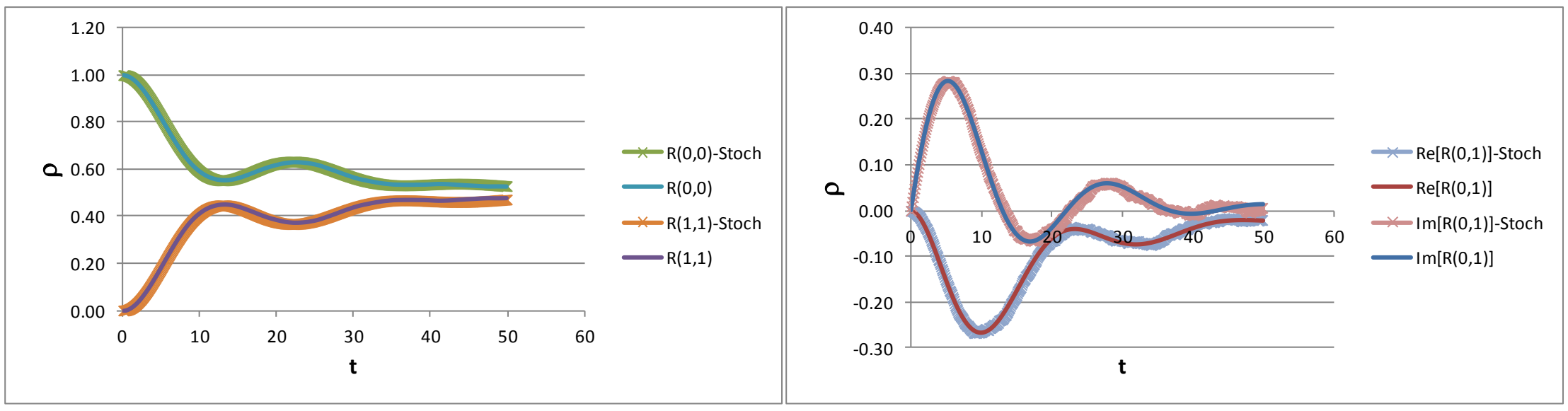

Figure 2.3: Same as Figure 2.1 but for $D = 0.1$.

### 2.4.1.4 Strong coupling to bath: $D = 0.3$.

The Lindblad constant of $D = 0.3$ is in the over-damped limit and no oscillations are observed while populations decay to 0.5 and coherences to zero as a single exponent with a time constant of $\lambda = 0.1$. The coherences seem to decay much faster, also we can see the deference between Stochastic where there are noise and non Stochastic which is smooth.

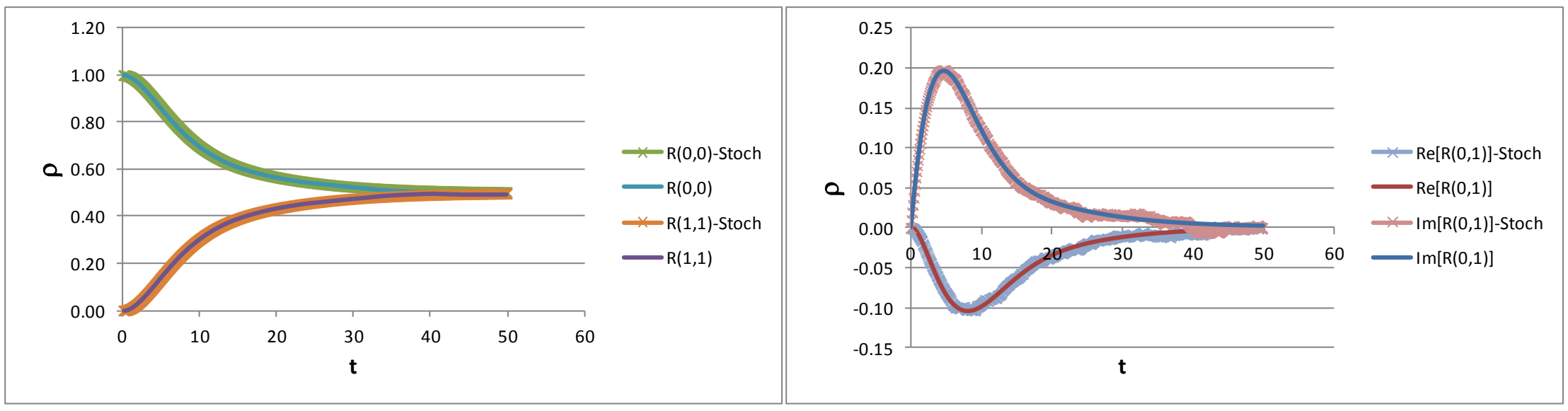

**Figure 2.4: Same as Figure 2.1 but for $D = 0.3$.**

### 2.4.2 Case of unequal parameters

#### 2.4.2.1 Medium coupling with $\epsilon = 0.2, \Delta = 0.4$

We now take a look at the case where the ratio between $\epsilon/\Delta$ is not 1 as above but 0.5. First we look at $D = 0.1$ i.e. the Lindblad equation reduces to a closed system Liouville-von-Neumann equation. The populations and coherences are as given below.

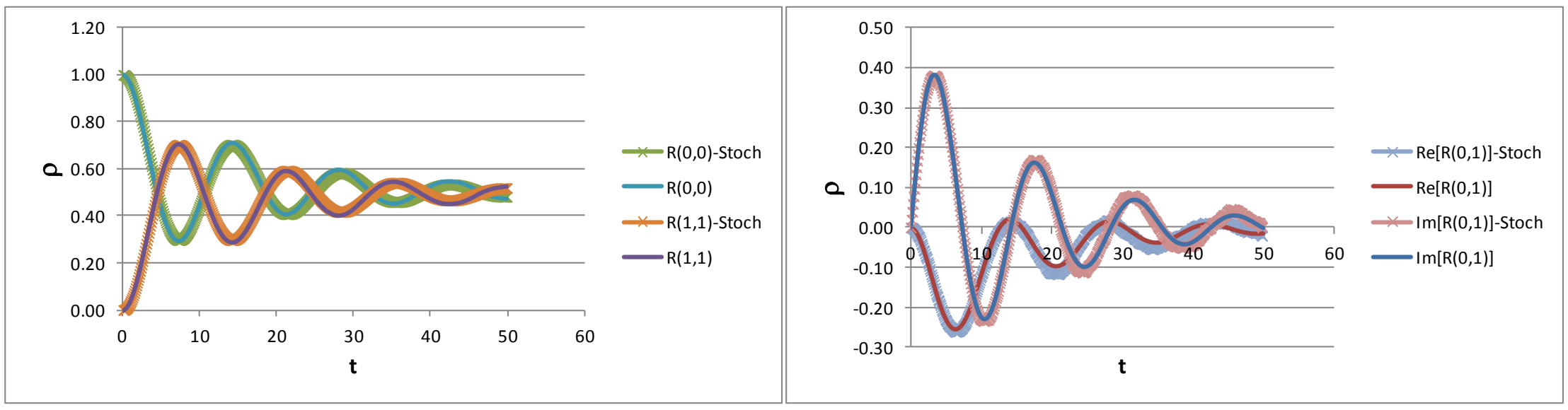

**Figure 2.5: Same as Figure 2.1 but for $\epsilon = 0.2, \Delta = 0.4$ and $D = 0.1$.**

#### 2.4.2.2 Medium coupling with $\epsilon = 0.2, \Delta = 0.1$

In this case we took $\epsilon/\Delta = 2$ and $D = 0.1$, we see that the decay take more time than the previous case which can explained by Rabi oscillation, also there are less oscillations comparing than previous one.

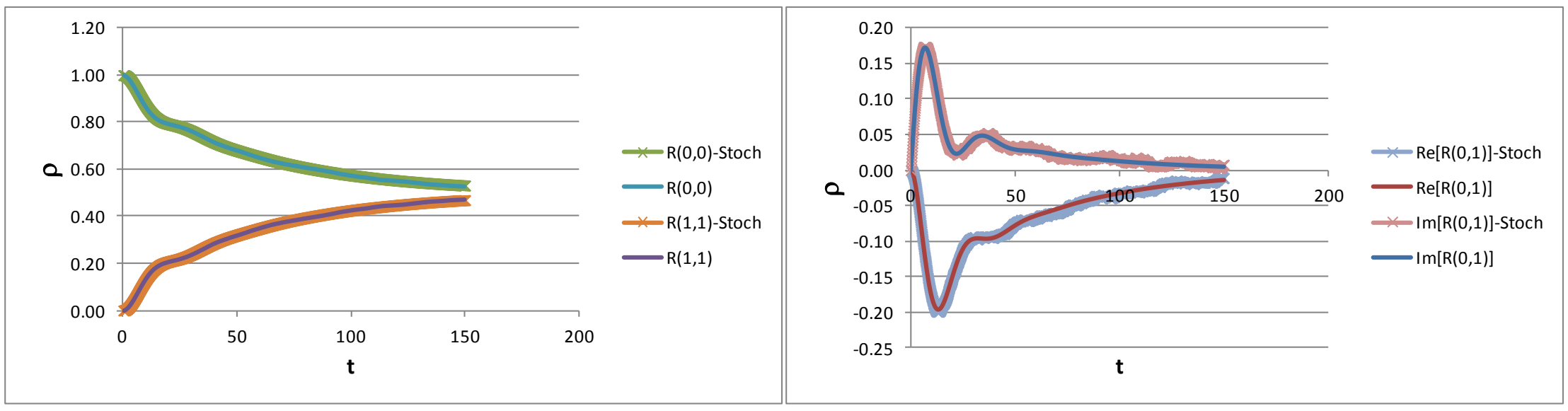

**Figure 2.6: Same as Figure 2.1 but for $\epsilon = 0.2, \Delta = 0.1$ and $D = 0.1$.**

We notice that the diagonal of the density matrix goes to half in the cases above, we explain that in Appendix 1.

### 2.4.3 Strong coupling, pure dephasing ($\Delta = 0$) from a mixed state

Here we study a case of pure dephasing, where $\Delta = 0$ where the operator $x$ in the Lindblad equation (2) is the unperturbed Hamiltonian $\hat{x} = \hat{H}_0$. Here we took $\epsilon = 0.2$.

In this example we start the dynamics from a *mixed state* (in all previous examples we started from a pure state):

$$\hat{\rho}_{t=0} = w_0 |\psi_0\rangle\langle\psi_0| + w_1 |\psi_1\rangle\langle\psi_1|$$

Where the eigenvectors of $\hat{\rho}$ are characterized by an angle $\alpha$:

$$|\psi_0\rangle = \begin{pmatrix} \cos\alpha \\ \sin\alpha \end{pmatrix}, |\psi_1\rangle = \begin{pmatrix} -\sin\alpha \\ \cos\alpha \end{pmatrix}$$

And we take $w_0 = 0.7, w_1 = 0.3, \alpha = \frac{\pi}{4}$. Thus:

$$\hat{\rho}_{t=0} = \begin{pmatrix} 1/2 & 1/5 \\ 1/5 & 1/2 \end{pmatrix}$$

For pure dephasing the population of each level (diagonal elements of $\rho$) is constant but the coherences (off diagonal elements) decay to zero so that in the infinite time limit the density matrix goes to a diagonal matrix:

$$\rho = \begin{pmatrix} 1/2 & 1/5 \\ 1/5 & 1/2 \end{pmatrix} \to \begin{pmatrix} 1/2 & 0 \\ 0 & 1/2 \end{pmatrix}$$

This behavior is indeed seen in Figure 2.7 the results shown below of the numerically exact and the stochastic Schrödinger equation approaches. It is seen that the time scale for the decay of the coherence is 7 time units.

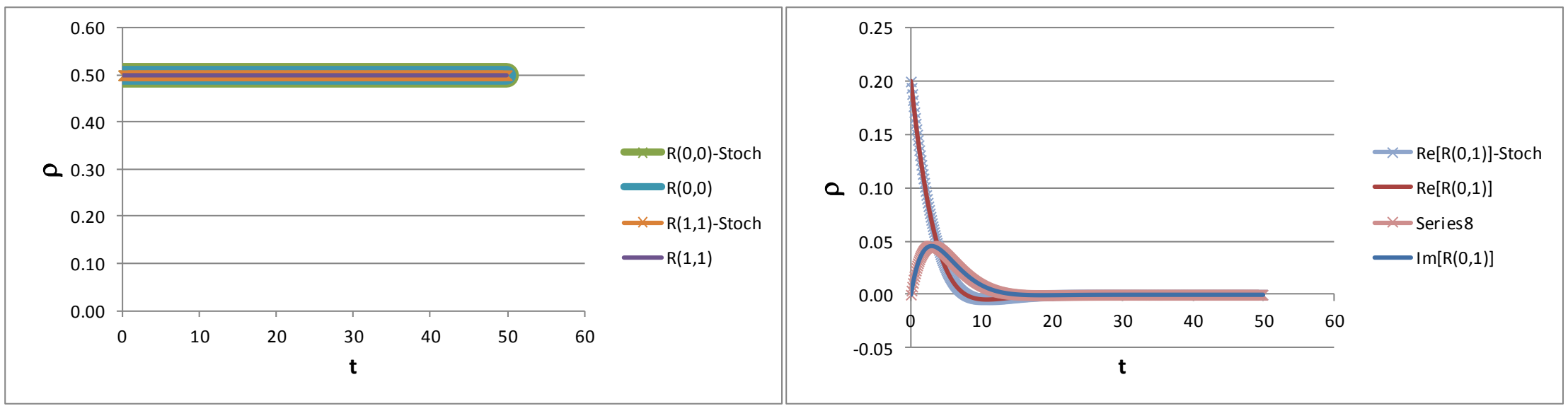

**Figure 2.7: Same as Figure 2.1 but for a mixed initial state and pure dephasing $\epsilon = 0.2, \Delta = 0$ and $D = 0.3$.**

## 2.5 Summary

In this chapter we studied the open 2-level system with the Lindblad approximation. We showed analytically that the time-dependent DM resulting from the TD Lindblad equation (Eq. (2)) is equivalent to the average of pure state solutions to the stochastic Schrödinger equation (Eq. (4)). We also gave numerical results which support this for the 2-level system case. In all cases shown above the stochastic dynamics yield results which are essentially identical to the numerically exact Lindblad dynamics.

The question now is whether the stochastic Schrödinger equation approach is also suitable for more complicated systems, like those in real space. This is discussed in the next chapter.

# 3 Quantum Langevin Equation Approach

## 3.1 A problem in using the stochastic Schrödinger equation

The experience with the two-level system gave us hope that the hermitian stochastic Schrödinger equation approaches will be useful for real-space systems like a particle in a potential well. Such systems are best treated numerically using grids.

For development purposes we consider a free particle interacting with a bath, where the bath operates on the particle through a harmonic mode:

$$H_0 = \frac{p^2}{2\mu} + \frac{1}{2}kq^2 \qquad (6)$$

The ground-state of the particle in the Harmonic potential is a Gaussian function $\psi_{gs}(q)$. Schrödinger equation propagation starting at $t = 0$ from unit-displaced wave function $\psi_{gs}(q+1)$ the position $\langle q \rangle_t$ as a function of time is identical to that of a classical harmonic oscillator with the same initial conditions: with the total energy is constant and the kinetic and potential energies oscillate as a function of time with frequency $\omega = 1$ (period $T = 2\pi$).

The fact that free particle interacts with a bath can be described either through a Lindblad term or through a stochastic Schrödinger equation as we did in the previous chapter for a 2 dimensional system. In addition to the Harmonic potential, the bath also fluctuates we introduce a stochastic force $f_m = -V_S'(m\Delta t)$ operating at each time-step $t_m - t_{m-1} = \Delta t$, $m = 1,2,\ldots$ by adding to the Hamiltonian a linear potential $V_s(t_m) = f_m q$ with

$$f_m = \frac{\sigma_f}{\sqrt{\Delta t}}\eta_m \qquad (7)$$

$\eta_m$ zero-mean unit-variance Gaussian variates and $\sigma_f$ the amplitude of the force fluctuations. Thus, the time dependent Hamiltonian is:

$$H(t_m) = H_0 + f_m\hat{q} \qquad (8)$$

As we show in Appendix 3 the Lindblad dynamics is obtained by averaging the dynamics of such a Hamiltonian when $f_m$ is a discretized white-noise random force with covariance (compare with the continuous version, Eq. (5)):

$$\langle f(t_m) \rangle = 0 \qquad (9)$$

$$\langle f(t_n)f(t_m)\rangle = \sigma_f^2 \times \frac{\delta_{nm}}{\Delta t}.$$

We plot the momentum expectation value as a function of time shown in the left panel of Figure 3.1. The momentum starts from zero then rises as part of an oscillation but due to the random force quickly decays and exhibits random fluctuations around $p = 0$. This is the expected behavior from a Harmonic oscillator connected to a dissipative (cold) bath.

However, examining the energies on the right-hand side of the figure shows an alarming behavior: the stochastic force heats up the system and the energy rises steeply to a non-physical value. This behavior is similar to that of a classical harmonic oscillator which under purely random force heats up. It is a quantum phenomenon that while $p$ goes to zero $p^2$ goes to “infinity”.

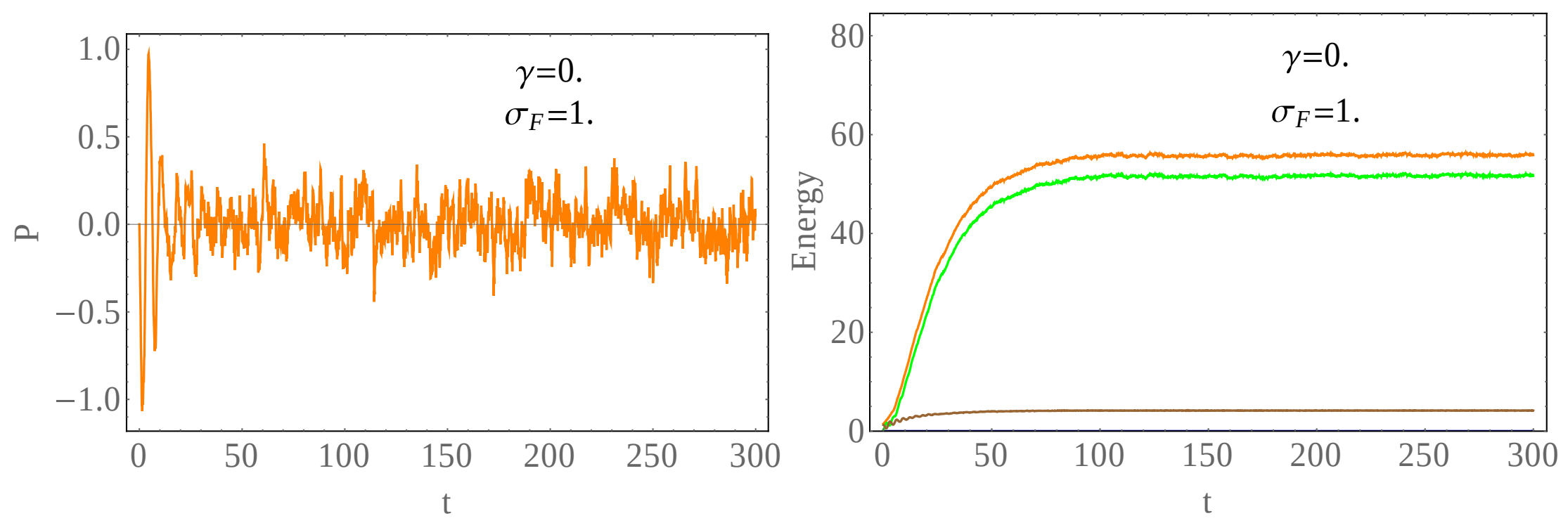

**Figure 3.1: Left panel: The momentum as a function of time starting the dynamics with a ground state displaced by a distance $x_0 = 1$. Right Panel: The total (orange), kinetic (green) and potential (brown) energies of the oscillator as a function of time starting from the displaced coherent state.**

It is a wonder that the stochastic Schrödinger equation which was shown in Appendix 3 to be equivalent *on the average* to the Lindblad equation give such numerically terrible results. We believe that this is due to the huge fluctuations that develop in the system, thus if we could get over the fluctuation we could show that on the average we obtain the physical result of the Lindblad equation. In short, the stochastic Schrödinger equation converges *on the average* but we are not able to obtain this average in practical calculations due to unmanageable fluctuations unless the system is a two level system, where the theory works beautifully, most likely because fluctuations are suppressed there.

## 3.2 Quantum Langevin Dynamics

In classical mechanics the remedy for eliminating the heating of the system (large fluctuations) is to introduce a *friction force*. [1] This can also be the solution for the quantum system and the Hamiltonian formulation of Zwanzig [10] and the quantum mean-field generalization of Peskin [11] serve to inspire a similar approach. In order to "Langevinize" our system we consider the time-dependent Hamiltonian where in addition to the stochastic force we enter a "deterministic" potential with an amplitude equal to the expectation value of the momentum:

$$H(q,p) = \frac{p^2}{2\mu} + \frac{1}{2}kq^2 + \gamma q\langle p\rangle + q\frac{\sigma_F}{\sqrt{\Delta t}}\eta \qquad (10)$$

Here, $p$ and $q$ are dynamical variables in classical mechanics or operators in quantum mechanics while $\langle p\rangle$ is the instantaneous momentum in the propagation. The Hamiltonin equations are:

$$\begin{aligned} \dot{p} &= -kq - \gamma\langle p\rangle - \frac{\sigma_F}{\sqrt{\Delta t}}\eta \\ \dot{q} &= p/\mu \end{aligned} \qquad (11)$$

In classical mechanics $\langle \mathrm{p}\rangle = \mathrm{p}$ so in actuality the first of these equations is:

$$\frac{d}{dt}\langle \mathrm{p}\rangle = -\mathrm{q} - \gamma\langle \mathrm{p}\rangle - \frac{\sigma_F}{\sqrt{\Delta \mathrm{t}}}\eta \qquad (12)$$

which is the Newton equation for a damped harmonic oscillator under random forces. In quantum mechanics, the Schrödinger equation is non-linear:

$$i\hbar\dot{\psi}(q,t) = \hat{H}_0\psi + \gamma\langle p\rangle\hat{q}\psi(q,t) + \frac{\sigma_F}{\sqrt{\Delta t}}\eta\hat{q}\psi(q,t) \qquad (13)$$

The effect of the "friction" term $\hat{\Gamma} = \gamma\langle p\rangle\hat{q}$ is to dissipate energy. Indeed, the energy of the non "friction" terms $E(t) = \langle\psi(t)|\hat{H} - \hat{\Gamma}|\psi\rangle \equiv \langle\hat{H} - \hat{\Gamma}\rangle_t$ must decay at a rate proportional to $\langle p\rangle^2$:

$$\dot{E}(t) = \frac{d}{dt}\langle\hat{H} - \hat{\Gamma}\rangle_t = -\frac{i}{\hbar}\gamma\langle p\rangle_t\langle[q,H]\rangle_t = -\frac{i}{\hbar}\gamma\langle p\rangle_t\left\langle-\frac{i\hbar p}{2\mu}\right\rangle_t = -\frac{\gamma}{2\mu}\langle p\rangle_t^2 \qquad (14)$$

Because the right hand side is negative the energy decays monotonically to zero. Neuhauser [12] has shown that when a term $\hat{J}\langle\hat{J}\rangle_t$ is added to a Schrödinger equation, where $\hat{J}$ is any Hermitian operator, it creates pure energy dissipation from the reset of the Hamiltonian. The case above is a special case of this theorem.

An analogous calculation of the energy $E_0 = \langle \hat{H}_0 \rangle_t$ will result in:

$$\dot{E}_0(t) = \frac{d}{dt}\langle \hat{H}_0 \rangle_t = -\frac{\gamma}{2\mu}\langle p \rangle^2 - \frac{\sigma_F}{2\mu\sqrt{\Delta t}}\eta\langle p \rangle \quad (15)$$

This does not necessarily decay to zero, however, since the statistical average of $\eta$ is zero, averaging on many trajectories will also result in a decay of the statistically averaged energy $E_0$ to zero. There remains the problem of the statistical fluctuations.

Finally, let us consider the expectation value of the momentum which will (by Ehrenfests theorem) obey Eq. (12). Averaging on many realizations will the random force contribution to $\frac{d}{dt}\langle p \rangle$ will be zero and the equation of motion would be $\frac{d}{dt}\langle \mathrm{p} \rangle = -\mathrm{q} - \gamma\langle \mathrm{p} \rangle$. Which is the damped Harmonic oscillator. Thus we expect the momentum to go to zero with a time constant of $\gamma^{-1}$. This is in accord with the fact that in thermal equilibrium the momentum expectation value is always zero:

$$\langle p \rangle_{k_B T} = tr\{\rho_{k_B T} p\} = 0, \quad (16)$$

because in any (bound) eigenstate of the Hamiltonian $H$ the momentum expectation value is zero:

$$\langle \psi_n | p | \psi_n \rangle = \mu \langle \psi_n | \dot{x} | \psi_n \rangle = \frac{i}{\hbar}\mu \langle \psi_n | [x, H] | \psi_n \rangle = \frac{i}{\hbar}\mu \langle \psi_n | x | \psi_n \rangle (E_n - E_n) = 0.$$

In classical mechanics the fluctuation-dissipation theorem (FDT) connects between the random force fluctuations the friction coefficient and the temperature (average kinetic energy) developed in the dynamics. In a time-discretized version this fluctuation-dissipation theorem becomes:

$$\langle f_n f_m \rangle \Delta t = k_B T \mu \gamma \delta_{nm} \quad (17)$$

Thus the force variance $\sigma_f^2 = \langle f^2 \rangle \Delta t$ must therefore obey:

$$\sigma_F^2 = k_B T \mu \gamma \quad (18)$$

We now describe some numerical results which are obtained by propagating the wave function using the Newton-Langevin Eq. (12) and Schrödinger-Langevin Eq. (13) for the vase of $k = 1$ and $m = 1$ (all in dimensionless units).

### 3.2.1 Numerical-dynamical details of the calculations

For the Newton-Langevin Eq. (12) we use Verlet propagation, this means that if at time $t^n$ we have the current position $r^n$ and velocity $v^n$ and we want to find the position and velocity in the next time step $t^{n+1} = t^n + dt$ of a particle of mass $\mu$ so we use the equations:

$$r^{n+1} = r^n + v^n dt + \frac{dt^2}{2\mu} f^n$$

$$v^{n+1} = v^n + \frac{dt}{2\mu}(f^n + f^{n+1})$$

Where $f^n = f(r^n, v^n)$ is the force at position $r^n$ and velocity $v^n$.

In our calculation we take small time step for high accuracy ($dt = 0.01$) and because we have a random force so we repeated the calculation $I = 1000$ times for each trajectory and averaging the results. According to the rules of probability theory the fluctuations in the averaged trajectory are $\sqrt{I}$ smaller than the fluctuations in a single trajectory.

Schrödinger-Langevin Eq. (13) is treated using a Fourier grid method, taking the grid to span the interval $[-5,5]$ of the $q$ axis, with $N_g = 56$ gridpoints. We used Chebyshev propagation [13] at time intervals of $\Delta t = 0.01$. To obtain the results equivalent to the Lindblad equation we averaged all expectation values over a repeated calculation, each time with a different random seed for the random force. The number of such repeated iterations was $I = 1000$.

### 3.2.2 Classical vs quantum dynamics without random force

We start with the classical case, taking as initial conditions

$$q_0 = 1, v_0 = 0.$$

The harmonic oscillator with no friction ($\gamma = 0$) is a closed system and there are no interaction with the environment. The oscillator oscillates indefinitely as shown in Figure 3.2.

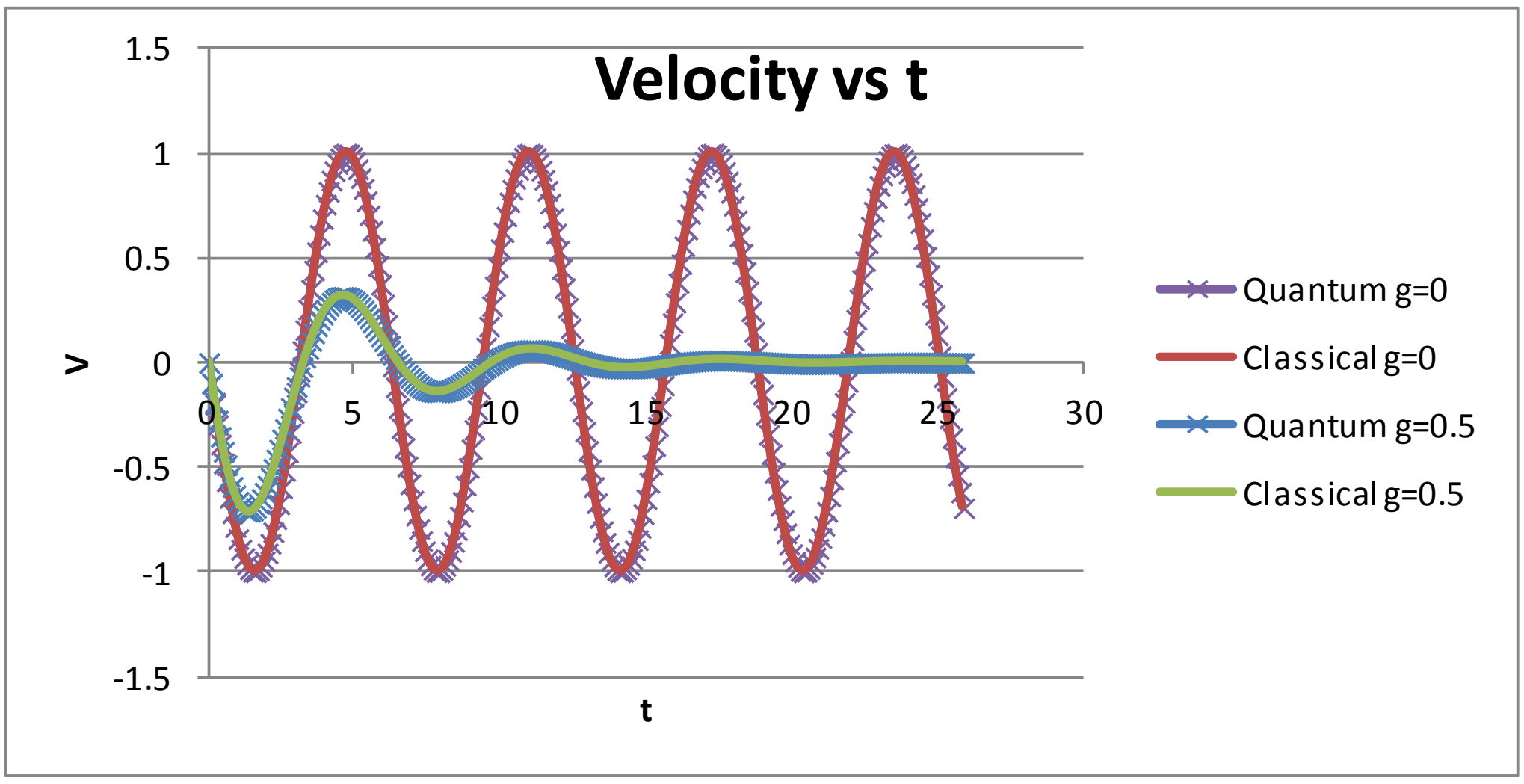

**Figure 3.2: Velocity vs. time for the damped/undamped Harmonic oscillator in the classical and quantum cases (see text)**

For $\gamma = 0.5$ the velocity of the harmonic oscillator decay to zero in time, within a time $\gamma^{-1}$, so the system is an open system which interacts with the environment. We can see also that $\gamma$ does not affect on the periodic time of the oscillator.

Now we consider the same problem but in quantum mechanics. We take as an initial state the displaced Gaussian $(q) = Ae^{-\frac{(q-q_0)^2}{2\sigma^2}}$, taking $q_0 = 1$ , $\sigma = 1$.

The results of this quantum propagation are also shown in Figure 3.2. The agreement with the classical results is perfect. This is due to the harmonic potential, in anharmonic cases there will not be such an exact agreement.

### 3.2.3 Classical vs quantum dynamics with random force

Starting from the same initial conditions as above we added a random force variance $\sigma_F^2$ taken from the fluctuation dissipation relation Eq. (18) with a temperature of $k_B T = 1$. The result of the classical run is shown in Figure 3.3. Here we can see that the kinetic energy oscillate until it reach the equilibrium where kinetic energy equal to $\frac{1}{2} k_B T$ in one dimension which is the classical definition of temperature.

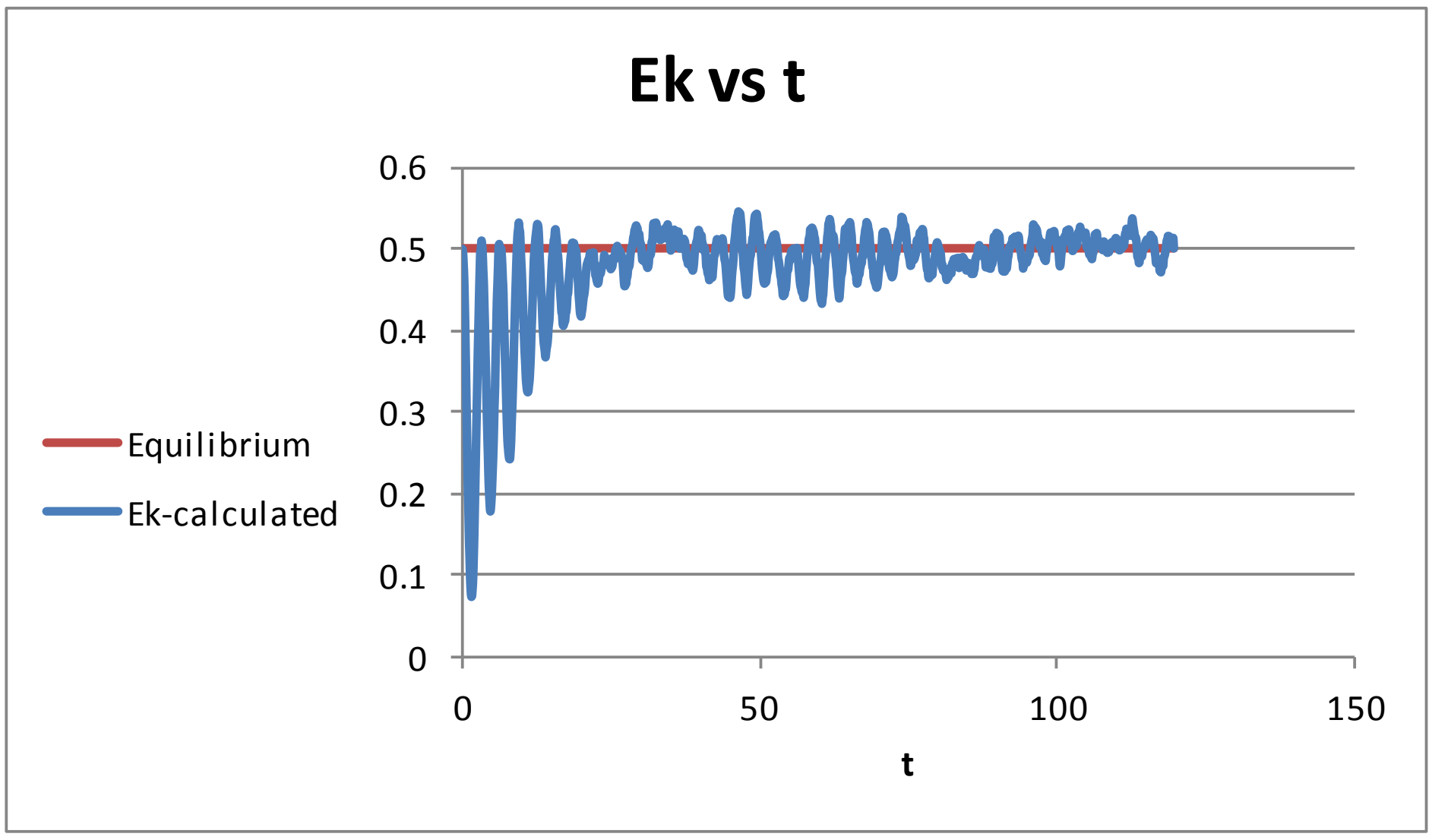

**Figure 3.3: The classical averaged kinetic energy as a function of time for the Langevin equation.**

There is no direct comparison of the figure with the quantum result. The reason is that in the quantum results there is always zero point kinetic energy, of at least $E_k = 0.25$ (half the ground state zero point energy). Thus, the effect of the temperature of the bath would be to add kinetic energy to the zero-point kinetic energy. This would preclude comparison. Furthermore as we show below, there instabilities in the nonlinear Schrödinger equation which do not allow us to heat the system, only to cool it. This means that the temperature of the bath must be lower than the energy of the initial state. We therefore stop the comparison to classical results and concentrate now only on the quantum results.

### 3.2.4 The effect of friction $\gamma$ on the quantum system

In Figure 3.5 we plot the position and velocity expectation values as a function of time for with friction and without, taking $\sigma_F = 0.14$ which, in the case of $\gamma = 0.1$ corresponds (by the FDT eq. (18)) to $k_B T = 0.1$ (i.e. $\sigma_F = 0.14$).

We first look at the energy in Figure 3.4. When $\gamma = 0$ the system absorbs energy from the random fluctuations and its energy rises almost monotonically. This problem was already alluded to in Figure 3.1. However, when $\gamma = 0.1$ in accordance with FDT, the energy decays to a finite value $E_\infty = E_{gs} + k_B T$ (in our case =0.6 energy units).

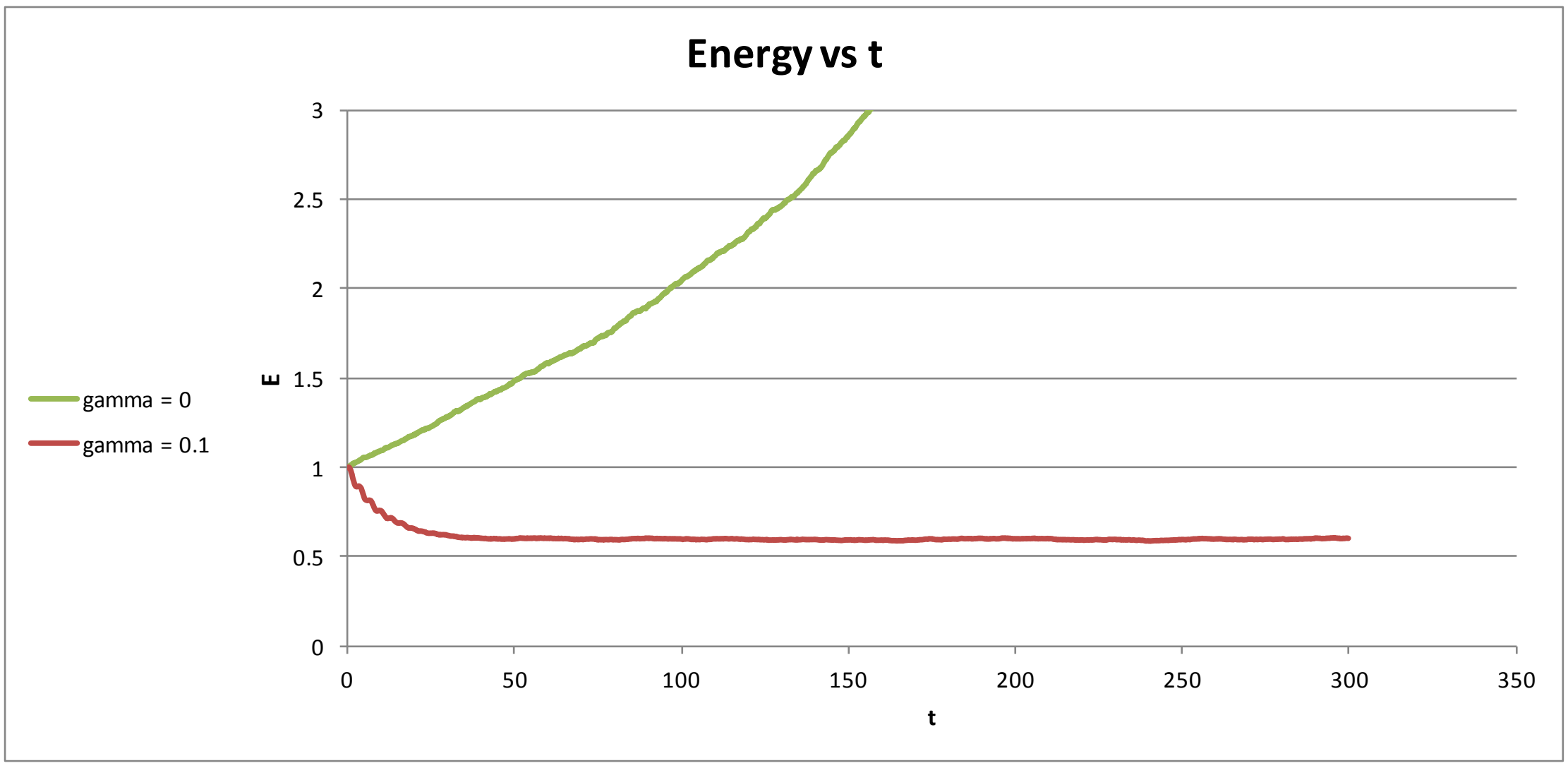

**Figure 3.4: The energy expectation value $\left\langle \frac{p^2}{2m} + \frac{kq^2}{2} \right\rangle_t$ as a function of time for zero friction and friction-fluctuation corresponding to a temperature of $k_B T$.**

Now we consider the position and velocity in Figure 3.5. For the $\gamma = 0.1$ we see that the initial velocities and position displacements oscillate but also decay with a time constant comparable to $\gamma^{-1}$ (it is not identical because of the fluctuations). The decay to zero is due to the fact that the as the system moves into a thermodynamical equilibrium the momentum must decay to zero (see Eq, (16)).

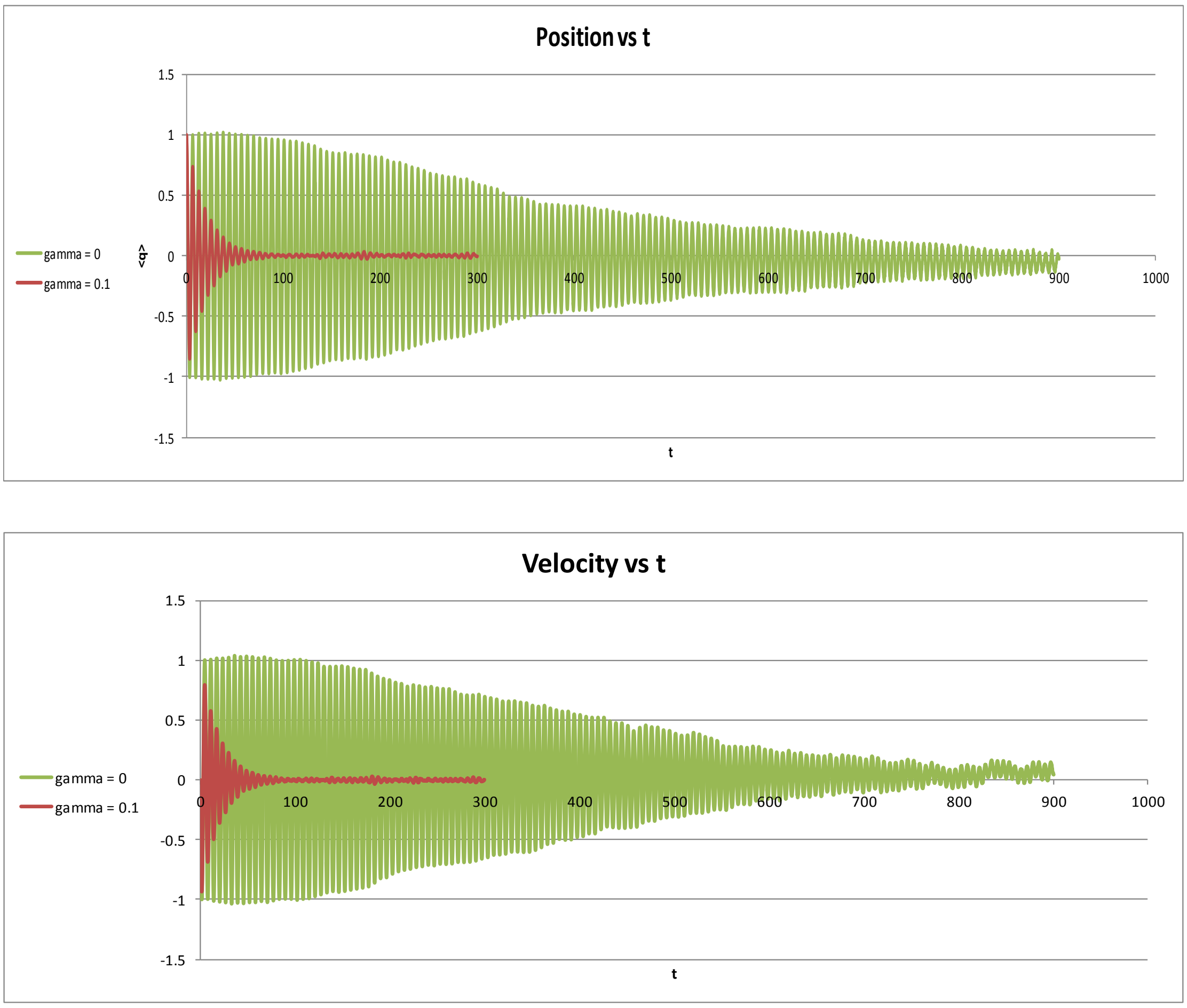

**Figure 3.5: The quantum dynamics of the position (top panel) and velocity (bottom panel) expectation values vs. time for $\gamma = 0$ and $\gamma = 0.1$.**

It is surprising that even when $\gamma = 0$ the velocity and position shown in Figure 3.5 decay to zero even though the energy grows (see Figure 3.4). This can be understood as a phenomenon of dephasing caused by averaging on many non-decaying random trajectories. Furthermore, in quantum mechanics, it is possible for $\langle p \rangle$ to go to zero while the kinetic energy $\langle p^2/2m \rangle$ to increase indefinitely, and same for $\langle x \rangle$ to decay to zero while the potential energy $\langle kx^2/2 \rangle$ to increase indefinitely.

### 3.2.5 Bath temperature effect on the system

Also we want to examine the effect of the bath temperature. This is done as follows. We first set $\gamma = 0.1$ and then for each temperature $T$ we determine through the FDT relation Eq. (18) the strength of the fluctuation $\sigma_F$. We plot the position and velocity as a function of time for several temperatures in Figure 3.6.

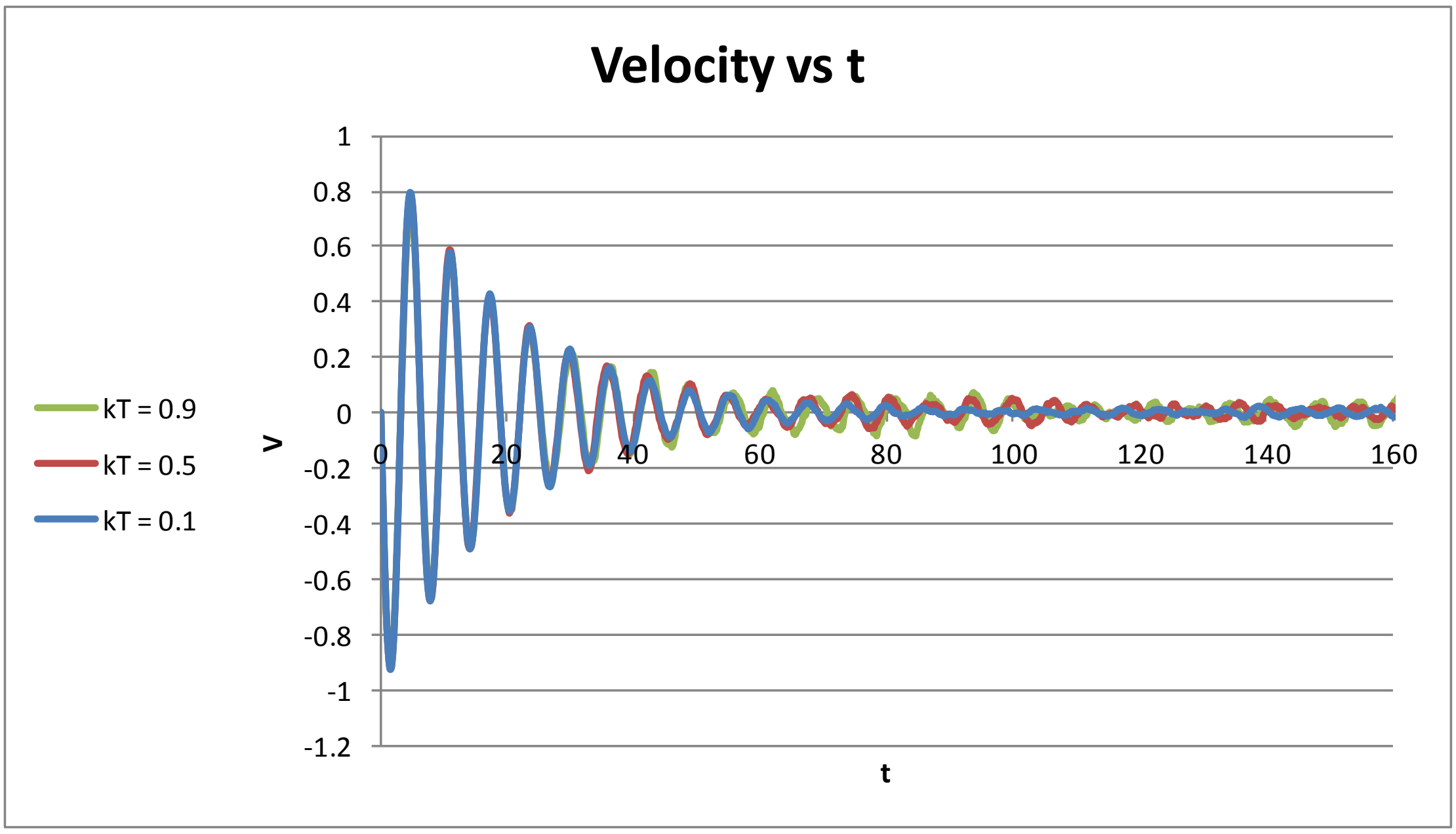

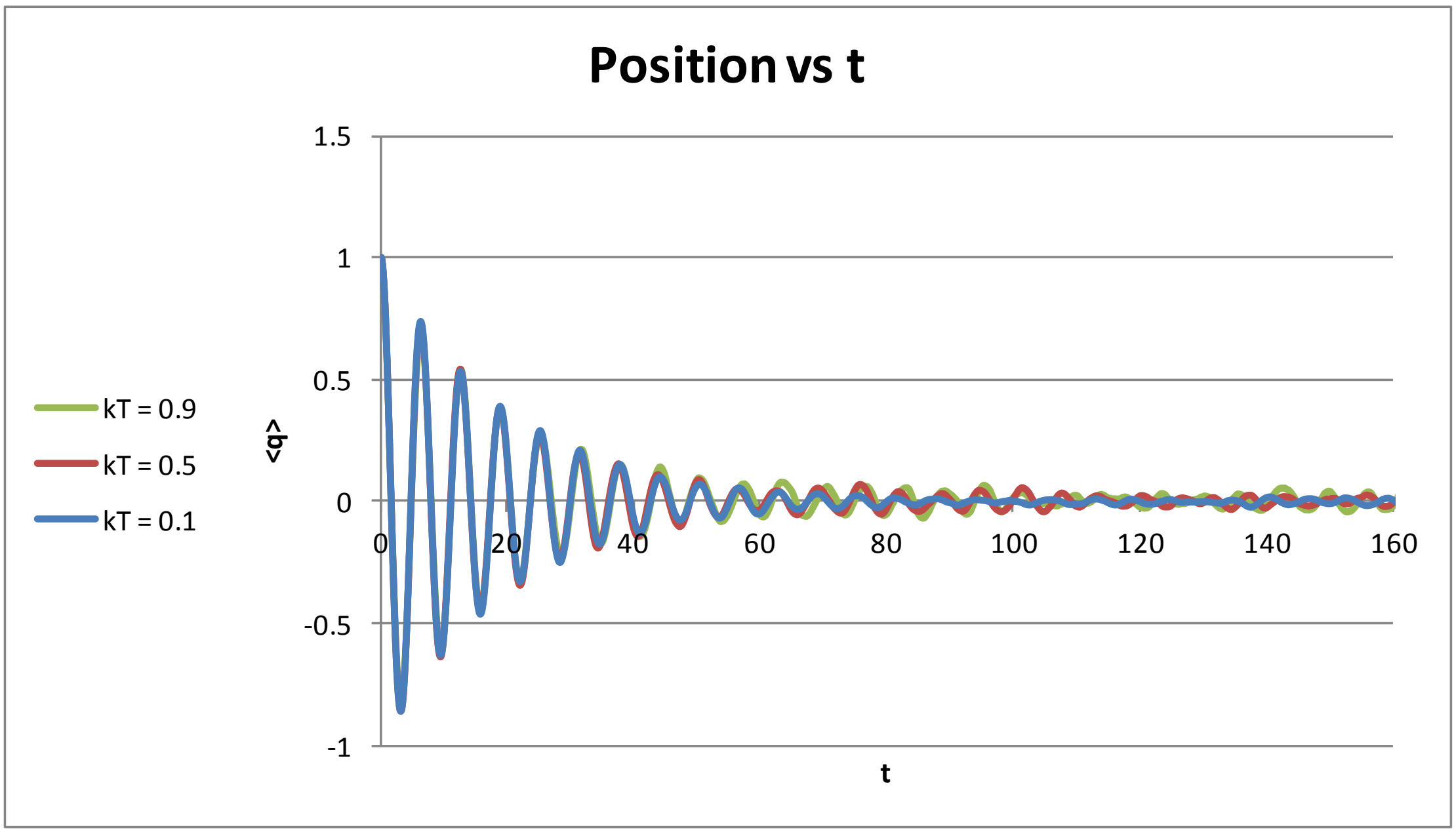

**Figure 3.6:the position and velocity expectation values as a function of time with deferent values of bath temperature of $k_B T$.**

We can see from Figure 3.6 that the expectation value of the position and velocity goes to zero in equilibrium. The dynamics is very similar and all three cases probably due to the fact that it is controlled by $k$ (the oscillations) and by $\gamma$ (the decay). However, if we study the energy, things look quite different as shown in Figure 3.7. Here we see that the bath temperature determined the final "equilibrium" energy of the system according to $E_{eq} = E_{g.s} + k_B T$. This is similar to the classical Langevin dynamics but with addition of the ground state energy, this is due to the fact that in quantum mechanics the energy of the system is the ground state energy even when for zero temperature so $k_B T = 0$.

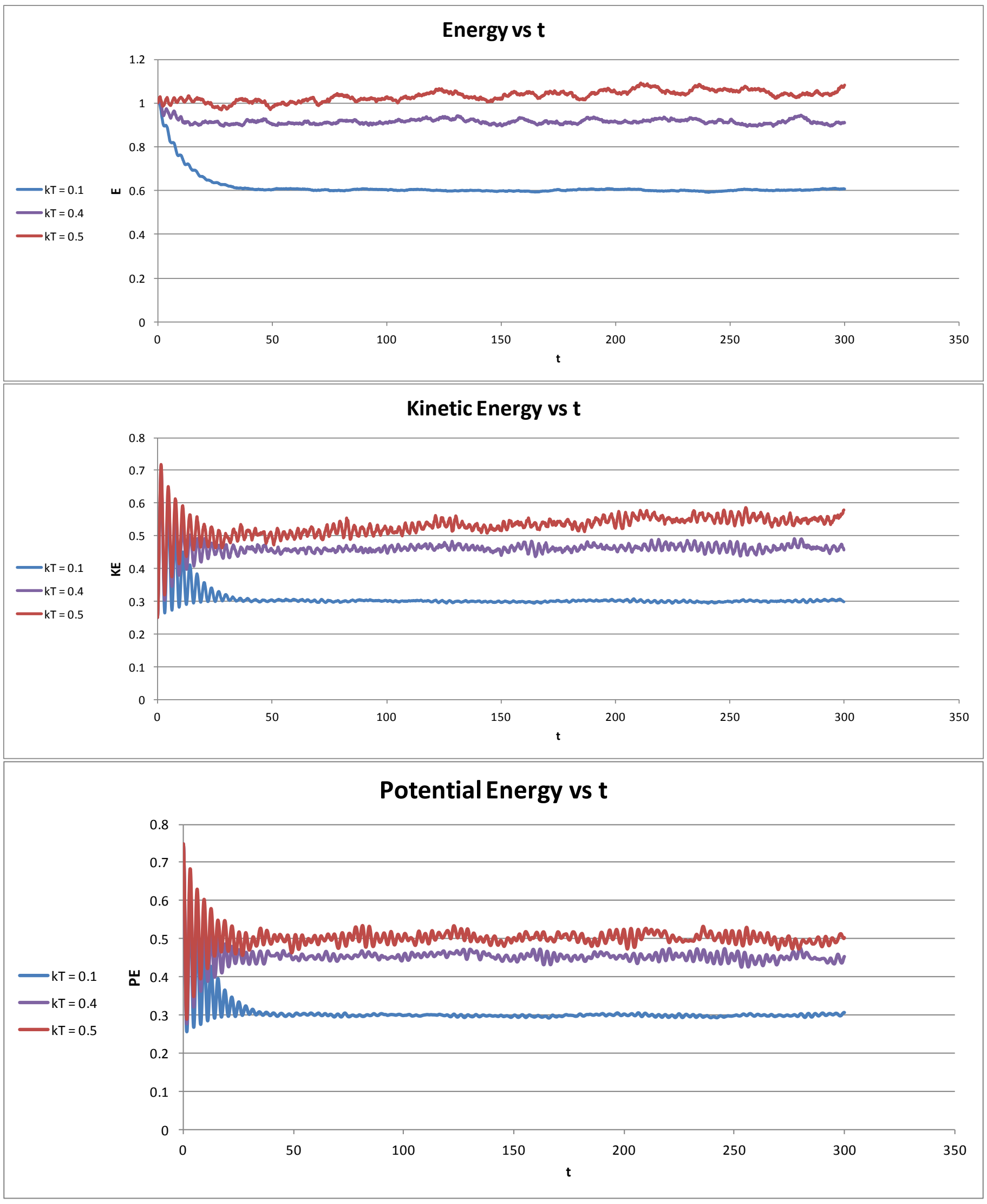

**Figure 3.7: The energy (top panel) and kinetic energy (middle panel) and potential energy (lower panel) expectation values as a function of time for different bath temperatures.**

We see in the middle panel of Figure 3.7 that the kinetic energy obeys the relation: $E_{k_{eq}} = E_{k_{g.s}} + \frac{1}{2}k_B T$ which the same relation in classical mechanics but with the ground state energy. The same happens for the potential energy in this case $V_{eq} = V_{g.s} + \frac{1}{2}k_B T$.

It can be noticed in the top and middle panels of Figure 3.7 that there is a problem in the graph of $k_B T = 0.5$: it seems that the energy grows slowly with time, i.e. that equilibrium is not attained. This problem becomes worse as the temperature increases, as shown in Figure 3.8 for $k_B T = 0.9$. The total energy and the kinetic energies seem to grow with time indefinitely once the temperature is higher than the ground state energy. We see from Figure 3.8 that the explosion in kinetic energy is faster than the potential energy, this is maybe due to the fact that the kinetic energy, being a second derivative of the wave function is more affected by the random fluctuations which cause the wave function to be non-smooth.

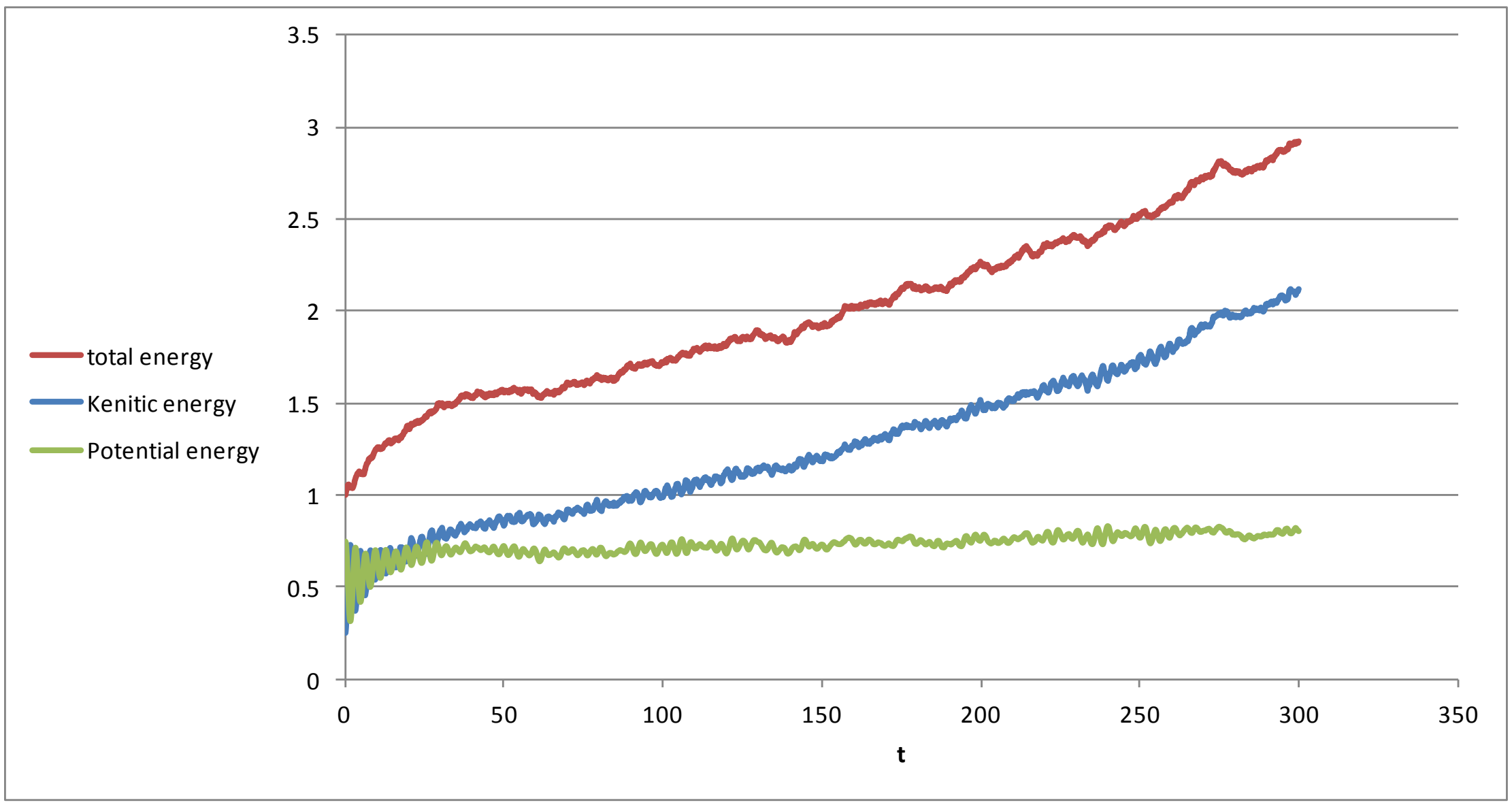

**Figure 3.8: Total kinetic and Potential energies expectation values as a function of time in a bath temperature of $k_B T = 0.9$**

# 4 Summary and Conclusions

Our goal in this work was to find a way to produce Lindblad dynamics without the need to use the density matrix. Instead we proposed to average over many quantum trajectories produced by a Hermitian Schrödinger propagation with a random force. This lead us to a stochastic quantum Schrödinger-Langevin dynamics which is similar to classical Langevin dynamics.

First we showed using a 2-level system that using the stochastic approach is equivalent to Liouville-von-Neumann approach by taking many examples, although the stochastic approach slower than density matrix approach in 2-level system but we hope the opposite dealing with large systems, like to take a large chain of 2-level systems then we can see the difference in the calculation.

After the 2-level system we were interested to apply the same idea to a system coupled to a Harmonic bath. The quantum system is given by a 2 one-dimensional particles interacting via a potential:

$$H_{\mathrm{S}} = \frac{p_1^2}{2\mu} + \frac{p_2^2}{2\mu} + V(q_1 - q_2) \tag{19}$$

and this system is connected to a Harmonic bath:

$$H_{\mathrm{B}} = T_B + V_B = \sum_{\mathrm{n}} \left( \frac{P_n^2}{2M_n} + \frac{1}{2} M_n \omega_n^2 Q_n^2 \right) \tag{20}$$

And the coupling $\Lambda_{\mathrm{SB}}$ represents a displacement of bath operators:

$$V_{\mathrm{B}} + \Lambda_{\mathrm{SB}} = \frac{1}{2} \sum_{\mathrm{n}} M_n \omega_n^2 \left[ \left( Q_n - \frac{\lambda_n}{2M_n\omega_n^2} q_1 \right)^2 + \left( Q_n - \frac{\lambda_n}{2M_n\omega_n^2} q_2 \right)^2 \right] \tag{21}$$

This model is due to Zwanzig [10] in classical mechanics. Peskin [11] generalized it for quantum mechanics using a mean-field approach. We discuss in Appendix 4 how the dynamics of the system can be mapped on a stochastic friction Schrödinger equation:

$$H(q,p) = H_S + \left[ \frac{1}{2} M\Omega^2 (q_1^2 + q_2^2) + \gamma [q_1 \langle p_1 \rangle + q_2 \langle p_2 \rangle] + \frac{\sigma_F}{\sqrt{\Delta t}} (\eta_1 q_1 + \eta_2 q_2) \right] \tag{22}$$

where $\sigma_F$ is given by fluctuation-dissipation relation and the spring constant by (see appendix 4):

$$\frac{1}{2}M\Omega^2 = \frac{1}{2}\mu\frac{\gamma}{\Delta t} \quad (23)$$

(here $\Delta t$ is the time step).

In this work we did not attempt to use this full model and only considered a single free particle connected to the bath. This prompted the damped-random-perturbed harmonic oscillator equation of Eq. (10).

When we did not have friction we found the random force heats the system up. This is not physical even though the expectation value of the momentum went to zero. Adding the friction term we could show that our approach exactly reproduces the classical dynamics in the absence of random force. Finally combining both friction and random force we saw the relation between the energy and the bath temperature which similar to the classical results, but we also noticed that the system can heat up if we give a higher temperature than the ground state energy.

We have also checked what happens when the coupling to bath is not Harmonic but through a Morse potential. Here we found that the dynamics is to heat up with out bounds at all temperatures. Thus, again, we find there are limits to the Langevin analogy.

Future research needs to address the problems and open questions we left from here:

1) Why does the system heat up, how can this be fixed?
2) Does the resulting dynamics really represent the Lindblad dynamics as we set out to achieve?
3) We have not used the relation in Eq. (23), and it is important to determine its effect.

# 5 Appendices

## 5.1 Appendix 1: Density matrix of two-level system in equilibrium

We can see that in equilibrium (which isn't pure dephasing) the density matrix gets the form:

$$\rho_{\mathrm{eq}} = \begin{pmatrix} \frac{1}{2} & 0 \\ 0 & \frac{1}{2} \end{pmatrix}$$

We can write the density matrix for two levels as:

$$\hat{\rho} = \frac{1}{2}\hat{\mathrm{I}} + a\hat{\sigma}_x + b\hat{\sigma}_y + c\hat{\sigma}_z$$

We can see that this equation obeys the condition: $tr(\rho) = 1$

So in equilibrium we know the fact that any operator which commute with Liouville-von-Neumann equation will not change, it's trivial. But other operators the expectation value for them should go to zero.

So in our case, just the identity operator commute with Liouville-von-Neumann equation which we can see in all dephasing graphs.

In the other hand if we don't take $\sigma_x$ in our equation so we have just $\sigma_z$ which commute with Liouville-von-Neumann equation so it doesn't changed and the ratio between $\rho_{00}$ and $\rho_{11}$ not like before, but also relate to $\sigma_z$, this called pure dephasing.

## 5.2 Appendix 2: Rabi solution for time-dependent 2-level system

We have two level system: $H_0 = \hbar\frac{\epsilon}{2}\sigma_z$ and $H_1 = \hbar\frac{\Delta}{2}\cos(\omega t)\,\sigma_x$ , and for simple calculation (setting $\hbar = 1$).

So $H_{tot} = H_0 + H_1$

$$H_{tot} = \hbar\begin{pmatrix} \frac{\epsilon}{2} & \frac{\Delta}{2}\cos(\omega t) \\ \frac{\Delta}{2}\cos(\omega t) & -\frac{\epsilon}{2} \end{pmatrix}$$

If we start at $t = 0$ in state $|1\rangle$ so we can write the population of state 1 and 2 to be $c_1(t)$ , $c_2(t)$ respectively:

$$|c_1(t)|^2 = \frac{\Delta^2}{\Omega_R^2} + \frac{\Omega^2}{\Omega_R^2}\cos^2\left(\frac{\Omega_R^2 t}{2}\right)$$

$$|c_2(t)|^2 = \frac{\Omega^2}{\Omega_R^2}\sin^2\left(\frac{\Omega_R^2 t}{2}\right)$$

Where $\Delta = \omega - \epsilon$ and $\Omega_R^2 \equiv \Omega^2 + \Delta^2$

## 5.3 Appendix 3: Derivation of the stochastic equations approach

In this appendix we will show that the Lindblad dynamics, derived from the equation:

$$\dot{\hat{\rho}}(t) = -\frac{i}{\hbar}[\hat{H}_0, \hat{\rho}(t)] - \frac{\sigma_f^2}{2\hbar^2}[\hat{q}, [\hat{q}, \hat{\rho}(t)]]$$

Where $\hat{q}$ is a Hermitian operator identical on the average to solving the time-dependent Schrödinger equation for a wave function $|\psi_n\rangle = |\psi(t_n)\rangle$ propagating under the system Hamiltonian with a discretized white-noise random force given in Eq. (9) and a (discrete time-dependent) Hamiltonian given in Eq. (8). The propagation of the wave function $|\psi_0\rangle$ according to the time-discretized Schrödinger equation is given by:

$$|\psi_\mathrm{n}\rangle = U(t_N, t_0)|\psi_0\rangle = \prod_{n=1}^{N} e^{-\frac{i}{\hbar}\hat{H}(t_{n-1})\Delta t}\,|\psi_0\rangle \approx \prod_{n=1}^{N} U_0(\Delta t)U_1(t_n, t_{n-1})\,|\psi_0\rangle$$

Where on the left side $U_0(\Delta t) = e^{-\frac{i}{\hbar}\hat{H}_0\Delta t}$, $U_1(t_n, t_{n-1}) = e^{-\frac{i}{\hbar}f(t_{n-1})\hat{x}\Delta t}$ and this represents an approximation to with error of order $O(\Delta t^2)$. We take $\Delta t$ to be small enough so that this error is negligible. Of similar accuracy would be to replace $U_1(t_n, t_{n-1})$ by it's Taylor's series accurate to third order:

$$|\psi_N\rangle \approx \prod_{n=\mathrm{N}}^{1} U_0(\Delta t)\left(1 - \frac{i}{\hbar}\Delta t\hat{x}f_n + \frac{1}{2}\left(-\frac{i}{\hbar}\Delta t f_n\right)^2 \hat{x}^2\right)|\psi_0\rangle$$

$$= U_0(\Delta t)\left(1 - \frac{i}{\hbar}\Delta t\hat{x}f_N + \frac{1}{2}\left(-\frac{i}{\hbar}\Delta t f_N\right)^2 \hat{x}^2\right)|\psi_{N-1}\rangle \equiv$$

$$= U_0(\Delta t)\left(1 - iS_N - \frac{1}{2}S_N^2\right)|\psi_{N-1}\rangle$$

Where for brevity of notation let us denote the dimensionless action:

$$S_N \equiv \frac{1}{\hbar}\Delta t f_N \hat{x}$$

The density matrix $\hat{\rho}(t_N)$ is obtained as an average of the projector:

$$\hat{\rho}(t_N) = \langle \quad |\psi_N\rangle\langle\psi_N| \quad \rangle_{\{f_1 \cdots f_N\}}.$$

And we can write therefor:

$$\hat{\rho}(t_N) = U_0(\Delta t)\left\langle \quad \left(1 - iS_N - \frac{1}{2}S_N^2\right)|\psi_{N-1}\rangle\langle\psi_{N-1}|\left(1 + iS_N - \frac{1}{2}S_N^2\right) \quad \right\rangle_{\{f_1 \cdots f_N\}} U_0^\dagger(\Delta t).$$

Because of the white noise nature of the random force the averaging can be split in time and so this equation is written as:

$$\hat{\rho}(t_N) = U_0(\Delta t)\left\langle\left(1 - iS_N - \frac{1}{2}S_N^2\right)\hat{\rho}(t_{N-1})\left(1 + iS_N - \frac{1}{2}S_N^2\right)\right\rangle_{\{f_N\}} U_0^\dagger(\Delta t).$$

Where the averaging is now only on $S_N$. The linear terms in $S_N$ are averaged to zero and when keeping only terms to order $\Delta t^2$ we obtain:

$$\hat{\rho}(t_N) = U_0(\Delta t)\left[\hat{\rho}(t_{N-1}) + \left\langle S_N\hat{\rho}(t_{N-1})S_N - \frac{1}{2}\left[S_N^2\hat{\rho}(t_{N-1}) + \hat{\rho}(t_{N-1})S_N^2\right]\right\rangle\right] U_0^\dagger(\Delta t)$$

Now we take the derivative with respect to $t_N = t_{N-1} + \Delta t$ and obtain the underlying equation for $\hat{\rho}(t)$. Remembering that $\dot{S}_N = \frac{1}{\hbar} f_N \hat{x}$ we find

$$\hat{\rho}(t_N) = -\frac{i}{\hbar}\left[\hat{H}_0, \hat{\rho}(t_N)\right] + \frac{1}{\hbar^2}\langle f_N^2\rangle\Delta t\, U_0(\Delta t)\left[2\hat{x}\hat{\rho}(t_{N-1})\hat{x} - \left[\hat{x}^2\hat{\rho}(t_{N-1}) + \hat{\rho}(t_{N-1})\hat{x}^2\right]\right]U_0^\dagger(\Delta t)$$

Using Eqs. (1) and (9) this leads directly to:

$$\dot{\hat{\rho}}(t) = -\frac{i}{\hbar}\left[\hat{H}_0, \hat{\rho}(t)\right] - \frac{\sigma_f^2}{\hbar^2}\left[\hat{x}, \left[\hat{x}, \hat{\rho}(t)\right]\right]$$

And is the same as Eq. (2) with the following substitution:

$$\frac{D}{2\hbar^2} = \frac{\sigma_f^2}{\hbar^2} \rightarrow \sigma_f = \sqrt{\frac{\mathrm{D}}{2}}$$

### 5.4 Appendix 4: Parameters of the the Quantum Langevin model

Following Peskin [11], we want to reduce the dynamics to a single effective bath mode of mass $M$ and frequency $\Omega$ which together with the random force represents the bath fluctuations and a friction force which represents dissipation. The target Hamiltonian is:

$$H(q,p) = H_S + \left[\frac{1}{2}M\Omega^2 q^2 + \gamma q\langle p\rangle + x\frac{\sigma_F}{\sqrt{\Delta t}}\eta\right] \tag{24}$$

Taking a discrete time propagation with time step $\Delta t$, the delta-functions in time become $\delta_D(t) = \theta(|t| - \Delta t)/\Delta t$. Thus, the force-force auto correlation is:

$$\langle F(t)F(t')\rangle = \frac{\sigma_F^2}{\Delta t}\delta_{tt'} = \frac{\sigma_F^2}{\Delta t}\theta(t - t' - \Delta t) = \sigma_F^2\delta_D(t - t') \tag{25}$$

And in order to obtain the FDT, $\langle F(t)F(t')\rangle = k_B T\mu\gamma\delta_D(t - t')$ ($\mu$ is the mass of the system particles) we demand:

$$\sigma_F^2 = k_B T\mu\gamma \tag{26}$$

From Peskin's paper:

$$\langle F(t)F(t')\rangle = k_B T\xi(t - t') \tag{27}$$

Where $\xi(t - t')$ is a sum over the harmonic bath fluctuations:

$$\begin{aligned} k_B T\xi(t - t') = \sum_j \lambda(\omega_j)^2 \Bigg( \left\langle q(\omega_j, 0)^2 \right\rangle \cos\omega_j t \cos\omega_j t' \\ + \frac{\left\langle p(\omega_j, 0)^2 \right\rangle}{m^2\omega_j^2} \sin\omega_j t \sin\omega_j t' \Bigg) \end{aligned} \tag{28}$$

We represent this sum using a density of states $\rho(\omega)$:

$$\begin{aligned} k_B T\xi(t - t') = \int \lambda(\omega)^2 \Bigg( \langle q(\omega, 0)^2 \rangle \cos\omega t \cos\omega t' \\ + \frac{\langle p(\omega, 0)^2 \rangle}{m^2\omega^2} \sin\omega t \sin\omega t' \Bigg) \rho(\omega) d\omega \end{aligned} \tag{29}$$

and assume:

$$\langle q(\omega,0)^2\rangle = \frac{\langle p(\omega,0)^2\rangle}{m^2\omega^2} = \frac{P^2}{m^2\omega^2} \tag{30}$$

Thus we obtain:

$$\begin{aligned} k_B T\xi(t-t') &= \frac{\overline{P^2}}{m^2}\int_{-\infty}^{\infty}(\cos\omega t\cos\omega t' + \sin\omega t\sin\omega t')\lambda(\omega)^2\frac{\rho(\omega)}{\omega^2}d\omega \\ &= \frac{\overline{P^2}}{m^2}\int_{-\infty}^{\infty}\cos\omega(t-t')\,\lambda(\omega)^2\frac{\rho(\omega)}{\omega^2}d\omega \\ &= \frac{k_B T}{m}\int_{-\infty}^{\infty}\cos\omega(t-t')\,\lambda(\omega)^2\frac{\rho(\omega)}{\omega^2}d\omega \end{aligned} \tag{31}$$

Now we take a density of states according to Debye:

$$\rho_D(\omega) = \frac{3}{2}\frac{\omega^2\theta(\omega_D - |\omega|)}{\omega_D^3} \tag{32}$$

With this the integral is:

$$\begin{aligned} \int_{-\infty}^{\infty}\cos\omega(t-t')\,\lambda(\omega)^2\frac{\rho(\omega)}{\omega^2}d\omega &\approx \frac{3}{2}\frac{\bar{\lambda}^2}{\omega_D^3}\int_{-\omega_D}^{\omega_D}\cos\omega(t-t')\,d\omega \\ &= \frac{3}{2}2\frac{\bar{\lambda}^2}{\omega_D^3}\left[\frac{\sin\omega_D(t-t')}{t-t'}\right] = 3\pi\frac{\bar{\lambda}^2}{\omega_D^3}\delta_D(t-t') \end{aligned}$$

And so:

$$k_B T\xi(t-t') = \frac{k_B T}{m}3\pi\frac{\bar{\lambda}^2}{\omega_D^3}\delta_D(t-t') \tag{33}$$

Combining Eqs. (25), (27) and (33) we find:

$$\sigma_F^2 = \frac{k_B T}{m}3\pi\frac{\bar{\lambda}^2}{\omega_D^3} \tag{34}$$

Combining this with the FDT Eq. (26) we find:

$$3\pi\frac{1}{m}\frac{\bar{\lambda}^2}{\omega_D^3} = \mu\gamma \tag{35}$$

Now consider the harmonic bath mode. From Peskin's paper we can show:

$$\frac{1}{2}M\Omega^2 = \frac{3}{2}\frac{\bar{\lambda}^2}{m\omega_D^2} = \frac{3}{2}2\pi\frac{1}{m}\frac{\bar{\lambda}^2}{\omega_D^3}\left(\frac{\omega_D}{2\pi}\right) = \left[3\pi\frac{1}{m}\frac{\bar{\lambda}^2}{\omega_D^3}\right]\left(\frac{\omega_D}{2\pi}\right) = \mu\gamma\left(\frac{\omega_D}{2\pi}\right)$$

Thus:

$$\Omega^2 = \frac{\mu}{M}\frac{\omega_D}{\pi}\gamma$$

We now identify the Debye cutoff frequency with our smallest time resolution:

$$\omega_D = \frac{\pi}{\Delta t}$$

We thus find the effective bath spring constant:

$$\frac{1}{2}M\Omega^2 = \frac{1}{2}\mu\frac{\gamma}{\Delta t} \tag{36}$$